\documentclass{aastex6}

\usepackage{epsfig}
\usepackage{hyperref}
\usepackage{graphicx}
\usepackage{epstopdf}
\usepackage{pslatex}
\usepackage{color}
\usepackage[caption=false]{subfig}
\usepackage{float}
\usepackage{subfloat}
\usepackage{cleveref}
 \usepackage{natbib}
 \usepackage{amsmath}
 \usepackage{longtable}

\newcounter{iso}
\newcommand{\rxn}{\refstepcounter{iso}%
                  (\oldstylenums{\theiso})}

\newcommand\osref[1]{(\oldstylenums{\ref{#1}})}

\defcitealias{2016ApJ...818...52T}{TTZ16}

\defcitealias{2017ApJ...843...33D}{DLT17}

\defcitealias{2018ApJ...853...18Z}{ZT18}

\begin{document}

\title{

The SOMA Radio Survey. I. Comprehensive SEDs of High-Mass Protostars from Infrared to Radio and the Emergence of Ionization Feedback} 
\bigskip\bigskip


\author{V. Rosero\altaffilmark{1,2,3}, K. E. I. Tanaka\altaffilmark{4,5,3}, J. C. Tan\altaffilmark{6,2}, J. Marvil\altaffilmark{1}, M. Liu\altaffilmark{2,3}, Y. Zhang\altaffilmark{7}, J. M. De Buizer\altaffilmark{8},  M. T. Beltr\'an\altaffilmark{9}}

\altaffiltext{1}{National Radio Astronomy Observatory, 1003 Lopezville Rd., Socorro, NM 87801, USA}
\altaffiltext{2}{Department of Astronomy, University of Virginia, Charlottesville, VA 22904, USA}
\altaffiltext{3}{Department of Astronomy, University of Florida, Gainesville, FL 32611, USA}
\altaffiltext{4}{Department of Earth and Space Science, Osaka University, Toyonaka, Osaka 560-0043, Japan}
\altaffiltext{5}{Chile Observatory, National Astronomical Observatory of Japan, Mitaka, Tokyo 181-8588, Japan}
\altaffiltext{6}{Dept. of Space, Earth and Environment, Chalmers University, SE-412 96 Gothenburg, Sweden}
\altaffiltext{7}{Star and Planet Formation Laboratory, RIKEN Cluster for Pioneering Research, Wako, Saitama 351--0198, Japan}
\altaffiltext{8}{SOFIA--USRA, NASA Ames Research Center, MS 232--12, Moffett Field, CA 94035, USA}
\altaffiltext{9}{INAF, Osservatorio Astrofisico di Arcetri, Largo E. Fermi 5, 50125 Firenze, Italy}

\begin{abstract}
We study centimeter continuum emission of eight high- and
intermediate-mass protostars that are part of the \emph{SOFIA} Massive
(SOMA) Star Formation Survey, thus building extended spectral energy
distributions (SEDs) from the radio to the infrared. We discuss the
morphology seen in the centimeter continuum images, which are mostly derived
from archival VLA data, and the relation to infrared morphology. We
use the SEDs to test new models of high-mass star formation including
radiative and disk-wind feedback and associated free-free and dust
continuum emission \citep*[]{2016ApJ...818...52T}. We show that interferometric data of the centimeter continuum flux densities provide additional, stringent tests of the models by constraining the ionizing luminosity of the source and help to break degeneracies encountered when modeling the infrared-only SEDs, especially for the protostellar mass. Our derived parameters are consistent with physical parameters estimated by other methods such as dynamical protostellar masses. We find a few examples of additional stellar sources in the vicinity of the high-mass protostars, which may be low-mass young stellar objects. However, the stellar multiplicity of the regions, at least as traced by radio continuum emission, appears to be relatively low.

\end{abstract}


\section{Introduction} \label{sec:intro}

High-mass stars ($m_* \geq 8 M_\odot$) are important throughout
astrophysics, but their mechanism of formation is still actively
debated (see, e.g., \citealp{2014prpl.conf..149T} for a review). For
traditional star formation models based on Core Accretion, there is a
proposed evolutionary sequence as the protostar grows in mass. For
example, based on the Turbulent Core Accretion model
\citep{2003ApJ...585..850M}, \citet*{2014ApJ...788..166Z} and \citet{2018ApJ...853...18Z}
have
presented a sequence of protostellar evolution and infrared continuum
radiative transfer models exclusively developed for intermediate and
high-mass stars. These can be compared to observed infrared and sub-mm
spectral energy distributions (SEDs) and images to constrain the
properties of the protostar
\citep[e.g.,][]{2013ApJ...766...86Z, 2017ApJ...843...33D}.

High-mass protostars are generally expected to become fairly bright
centimeter continuum sources (flux densities of $\sim$ few mJy to Jy),
as the stellar photosphere heats up and begins to ionize its
surroundings.  Centimeter continuum observations, especially with
the improved capabilities of the Karl G. Jansky Very Large Array
(VLA)\footnote{The National Radio Astronomy Observatory is a facility
  of the National Science Foundation operated under cooperative
  agreement by Associated Universities, Inc.}, are therefore able to provide
unique insights into the earliest, embedded phases of high-mass star
birth. Moreover, \citet*[][hereafter TTZ16]{2016ApJ...818...52T}
calculated the predicted ionization structures and centimeter
continuum emission properties using the initial parameters resulting
from the radiative transfer models. In this framework, the earliest
stages of ionizing feedback involve the ionization of a
magnetohydrodynamical driven disk wind and/or X-wind, which would
appear as a thermal radio jet \citep[see also][]{2003astro.ph..9139T}.
Later, once the outflow is fully ionized, the ionizing photons begin
to interact with the infall envelope and disk, potentially driving a
photoevaporative outflow. Alternative models have been discussed in
the literature, including those involving ionized accretion flows at
the center of the core \citep{2007ApJ...666..976K}. More radically,
alternative formation scenarios invoking competitive accretion
\citep{2001MNRAS.323..785B, 2010ApJ...709...27W} or even protostellar
collisions \citep{1998MNRAS.298...93B, 2005AJ....129.2281B} are
expected to involve much more disordered accretion flows to the
protostar (see also \citealt{2005MNRAS.358..291D}) and these may
become illuminated by their radio emission once the protostars start
ionizing their surroundings.

High-mass star-forming regions
are composed of one or more
cores or dense substructures occupying $\lesssim$0.1~pc scales. When
there are multiple sources present, these  may be in different
evolutionary stages, from prestellar to protostellar. The latter begin
to show greater astrochemical complexity as the protostar warms up the
inner region, appearing as hot molecular cores (HMCs).
\citet{2016ApJS..227...25R}, using the VLA and achieving image rms
noise values of $\sim$3--10 $\mu$Jy/beam, found that HMCs are very
commonly associated with centimeter wavelength sources, most of them
with low radio emission levels on the order of ${\scriptstyle <}\,$1
mJy.  Moreover, many of these centimeter continuum sources have
morphologies and parameters that resemble ionized jets, which is in
general agreement with results from the
\citetalias{2016ApJ...818...52T} modeling. Thus, radio continuum
emission is extremely relevant in order to constrain the ionizing
luminosity of the protostar.

Physical parameters such as the mass of the hosting core, the mass of
the central protostar and the mass accretion rate are extremely
important to characterize the evolution of forming high-mass stars, but
obtaining accurate estimates of these parameters is rather
challenging. Thus, the testing and calibration of new theoretical models
which predict such parameters using observational data are urgently
needed.   Our overall goal now is to assemble multi-wavelength data for
a statistically significant sample of high- and intermediate-mass
protostars and use the data to test theoretical models of their
formation and feedback mechanisms. Our sample will probe
different environments, evolutionary stages and core masses using
observations from the MIR, FIR, sub-mm/mm and centimeter wavelengths.
The \emph{SOFIA} Massive (SOMA) Star Formation Survey (PI: Tan) aims
to observe $\sim$50 high- and intermediate-mass protostars with
\emph{SOFIA}-FORCAST at $\sim$10--40~$\mu$m. Such data, together with
ancillary {\it Herschel} data, help define the peak of the SED, thus
constraining the bolometric luminosity of the protostars. Furthermore
the $\sim$10--40 $\mu$m images trace warm dust delineating
protostellar outflow cavities, which helps determine the geometry of
the protostellar source. Extinction of this cavity emission by the
colder, dense core envelope also constrains properties of the
protostellar core.  The SOMA survey sample consists of a range of
 four source types  that have been defined  in \citet{2017ApJ...843...33D} (i.e., Type I to Type IV), from isolated MIR sources within otherwise
infrared dark clouds (IRDCs) lacking centimeter continuum emission
(i.e., Type I) at an image rms level $\sim$0.3 mJy/beam (typical of
the CORNISH survey; \citealt{2008ASPC..387..389P}), to more evolved
sources with known centimeter continuum emission of various extents
(i.e., Type II and Type III for association with hypercompact(HC) or
ultracompact(UC) HII regions, respectively), to clustered (within
$\sim$$60^{\prime \prime}$) MIR sources that are sometimes known
to be associated with radio emission (i.e., Type IV). So far, 22
sources from the SOMA survey have been observed with
\emph{SOFIA}-FORCAST and results for the first eight protostars from
the sample are presented in \citet[][hereafter
  DLT17]{2017ApJ...843...33D}. These results include the derivation of
SEDs and then model fitting using the theoretical \citet*[][hereafter
  ZT18]{2018ApJ...853...18Z} models developed exclusively for high-
and intermediate-mass protostars. In particular, the protostellar
mass, accretion rate and core envelope mass are estimated. The
geometries of the cores are also constrained: e.g., the near-facing,
blue-shifted side of the outflow has a cavity that typically appears
brighter at shorter IR wavelengths compared to the far-facing,
redshifted side. However, there are still significant degeneracies in
parameters derived from infrared SEDs alone, such as the protostellar
masses and bolometric luminosities that even for the best models have
relatively large allowed ranges. Our goal with this study is to use
the centimeter emission data and the associated predictions of
centimeter continuum free-free emission from
\citetalias{2016ApJ...818...52T} to help break these degeneracies.

Previous observational studies such as the Red MSX Source (RMS) survey
  \citep[e.g.,][]{2005IAUS..227..370H, 2007A&A...476.1019M,
    2009A&A...501..539U} have made progress towards multi-wavelength
  observations and studies of high-mass protostars located throughout
  the Galaxy. Their observations are mainly based upon the Midcourse
  Space Experiment (MSX) survey \citep{2001AJ....121.2819P} and Two
  Micron All Sky Survey (2MASS) data for the infrared as well as
  centimeter continuum from the VLA at 6 cm with spatial resolutions
  of $\sim 1-2^{\prime \prime}$ and image rms noise of $\sim0.22$
  mJy. However, our FORCAST data has $\sim 6 \times$ higher resolution
  that the MSX survey at $\sim 20\,\mu$m, and we are often resolving
  multiple sources of emission that appear as single sources in MSX,
  allowing for more precise photometry. Also, with the FORCAST data we
  have extended the wavelength coverage of the SEDs beyond the longest
  filter of MSX, which was $21.3\,\mu$m. Furthermore, the FORCAST
  filters have narrower bandpasses, and are therefore more accurate at
  estimating the flux at a given wavelength.

Previous relevant theoretical studies have included numerical
  simulations of massive star formation that incorporate MHD outflow
  feedback \citep[e.g.,][]{2011ApJ...740..107C, 2016ApJ...832...40K, 2017MNRAS.470.1026M, 2018arXiv181100954S, 2018A&A...616A.101K, 2018A&A...620A.182K}. However, in general the outputs of these simulations have
  not been coupled to detailed radiative transfer calculations for
  both the thermal dust and radio continuum (free-free) emission from
  ionized gas. Nor do these models span a wide range of environmental
  conditions. Thus our approach in this paper is to use the
  observational data to test simpler semi-analytic models, i.e., the
  ZT18 grid of models for thermal dust emission from massive
  protostars and the TTZ16 models for radio continuum emission from
  ionized gas calculated self-consistently from the ZT18 physical
  models.
%


In this paper, we present extended SEDs to include centimeter
continuum fluxes for the eight regions studied by
\citetalias{2017ApJ...843...33D} and which are under the SOMA Type II
category, specifically because the MIR emission extends beyond the
observed radio emission. All these sources are known to be associated
with large scale molecular outflows except for IRAS 07299$-$1651 due to
its limited observational data. We primarily used high-sensitivity Jansky
VLA public archival data at 6, 1.3 and 0.7 cm to determine the flux
densities of the regions in order to test the
\citetalias{2016ApJ...818...52T} model. We used information from the
literature when the regions did not have any public Jansky VLA data
available.  In order to investigate the morphology and the
multiplicity of the radio sources associated to our eight SOMA
regions, we focused on available high angular resolution data; in most
cases the data presented in this paper at the shorter wavelengths are
$\sim$10$\times$ higher resolution than at larger wavelengths. 
The
methodology and public archival data are presented in
\S\ref{methods}. A description of the \citetalias{2016ApJ...818...52T}
models is in \S\ref{Tanaka_models}. Basic observational results are
presented in \S\ref{results}, while analysis and testing of the
\citetalias{2016ApJ...818...52T} models are in \S\ref{sec:models}. The
discussion and summary are presented in \S\ref{discussion} and
\S\ref{summary}, respectively.

\section{Methods}\label{methods}

The SOMA Star Formation Survey sample is defined by {\it
  SOFIA}-FORCAST observations (i.e., from $\sim$10 to 40~$\rm \mu m$),
with the first eight sources presented by \citetalias{2017ApJ...843...33D}. These eight sources
define the sample for which we present and analyze the radio data in
this paper. The radio observations presented here are mostly public Jansky
VLA data retrieved from the VLA data archive, except for regions
G45.47$+$0.05 (at 6 cm) and Cepheus A where we used information
available in the literature and region IRAS 07299$-$1651 where we
present our own observations.

The archival and literature data that we analyzed in this work are
summarized in Table \ref{Observations_table}. Column 1 gives the
region name, and column 2, 3 and 4 give the band frequency, R.A., and
declination, while columns 5 and 6 give the synthesized beam size and
position angle (PA) and the rms of the resulting images. The distance to
every region, as adopted by \citetalias{2017ApJ...843...33D}, as well as the bolometric
luminosities evaluated by \citetalias{2017ApJ...843...33D}, are shown in columns 7 and
8, respectively.

The references for the radio fluxes that were obtained from the
literature are in column 9. A list of phase calibrators used in
the observations at 6 , 1.3 and 0.7 cm is given in Table
\ref{VLA_cal}.


\begin{deluxetable}{l c c c c c c c c}
\tabletypesize{\scriptsize}
\tablecaption{SOMA Sources:  Radio Continuum Data \label{Observations_table}}
\tablewidth{0pt}
\tablehead{
\colhead{Region}                  & 
\colhead{Frequency Band}  &
\colhead{R.A.}   &
\colhead{Dec. }               & 
\colhead{Beam Size}  & 
\colhead{rms}             & 
\colhead{D\tablenotemark{a}} & 
\colhead{L\tablenotemark{b}} & 
\colhead{Ref.\tablenotemark{c}} 
 \\[2pt]
\colhead{}                            & 
\colhead{(GHz)}                      & 
\colhead{(J2000)}                      & 
\colhead{(J2000)}                      & 
\colhead{($^{\prime \prime}$$\times$$^{\prime \prime}$, deg)}                      & 
\colhead{($\mu$Jy beam$^{-1}$)}                      &  
\colhead{(kpc)}                      & 
\colhead{(10${^4} L_{\odot}$)}                      & 
\colhead{}                         \\[-20pt]\\}
\startdata
\setcounter{iso}{0}	
AFGL 4029              & 4.0$-$8.0    & 03 01 31.28 & $+$60 29 12.9    &    0.35$\times$0.28, $-$43.7      & 7.0      & 2.0      & 1.6$-$34.0 & \nodata   \\[3pt]  
		       & 40.0$-$50.0    &  \nodata    &  \nodata         &    0.06$\times$0.04, $+$69.7      & 60.0      & \nodata  & \nodata & \nodata                                  \\[9pt] 	

AFGL 437               & 4.0$-$8.0    & 03 07 24.55 & $+$58 30 52.8    &    0.44$\times$0.33, $-$76.0      & 7.0      & 2.0      & 1.7$-$15.0 &\nodata   \\[3pt]  
		       & 40.0$-$50.0    &   \nodata    &  \nodata        &    0.06$\times$0.04, $+$71.7      & 60.0      &   \nodata   & \nodata & \nodata                                  \\[9pt] 

IRAS 07299$-$1651      & 4.0$-$8.0    & 07 32 09.74 & $-$16 58 11.3    & 0.47$\times$0.29, $-$9.84  & 7.0     & 1.68                   & 1.0$-$4.2  &\nodata      \\[3pt]  
		       & \nodata    & \nodata  & \nodata    & \nodata  &  \nodata    & \nodata  & \nodata & \nodata                                  \\[9pt] 

G35.20$-$0.74          & 4.0$-$8.0    & 18 58 13.02 & $+$01 40 36.2    &    0.54$\times$0.28, $-$48.5      & 8.0      & 2.2      & 3.8$-$8.4  &\nodata   \\[3pt] 
		       & 18.0$-$26.5    &   \nodata    &  \nodata        &    0.30$\times$0.25, $-$12.8      & 18.0      &   \nodata   & \nodata & \nodata                                  \\[3pt]  
		       & 40.0$-$50.0    &   \nodata    &  \nodata        &    0.17$\times$0.13, $-$29.1      & 15.0      &   \nodata   & \nodata & \nodata                                 \\[9pt] 

G45.47$+$0.05          & 6 cm\tablenotemark{d}    & 19 14 25.67 & $+$11 09 25.4    &    1.75$\times$1.57, $-$78.9                 & 1000.0      & 8.4      & 17.0$-$51.0 &\rxn \label{rxn:Urquhart2009}   \\[3pt]  
		       & 40.0$-$50.0    &  \nodata    &  \nodata         &    0.05$\times$0.04, $-$24.2      & 90.0      & \nodata & \nodata  & \nodata      \\[9pt] 

IRAS 20126$+$4104      & 4.0$-$8.0    & 20 14 26.05 & $+$41 13 32.5    &    0.33$\times$0.29, $+$65.2      & 6.0      & 1.64                   & 2.0$-$9.3 &\rxn \label{rxn:Rosero2016}      \\[3pt]  
		       & 18.0$-$26.5    &   \nodata    &  \nodata        &    0.35$\times$0.24, $-$85.1      & 10.0  & \nodata    &   \nodata     & \nodata                                \\[9pt] 

Cepheus A              & 3.6 cm\tablenotemark{d}    & 22 56 17.98  & $+$62 01 49.4    &    0.27$\times$0.19, $-$79.1      & 50.0      & 0.7                   & 2.4$-$9.9 & \rxn \label{rxn:Curiel2006}      \\[3pt]  
		       & 1.3 cm\tablenotemark{d}    &   \nodata    &  \nodata        &   0.09$\times$0.07, $+$32.8        & 50.0      &   \nodata     & \nodata    & \nodata                            \\[3pt] 
                       & 0.7 cm\tablenotemark{d}   & \nodata      & \nodata          &    0.05$\times$0.04, $-$57.9      & 250.0      &   \nodata     & \nodata            & \nodata                  \\[9pt]

NGC 7538 IRS9          & 4.0$-$8.0    & 23 14 01.77 & $+$61 27 19.8    &    0.32$\times$0.26, $+$20.7      & 30.0      & 2.65      & 3.7$-$8.2 &\nodata   \\[3pt]  
		       & 40.0$-$50.0    &  \nodata    &  \nodata         &    0.05$\times$0.04, $-$5.29      & 43.0      & \nodata              & \nodata & \nodata                      \\[9pt] 	
\enddata
\tablenotetext{\text{a}}{References cited in \citet{2017ApJ...843...33D}.}
\tablenotetext{\text{b}}{Range of bolometric luminosities from the best models reported in \citet{2017ApJ...843...33D}.}
\tablenotetext{\text{c}}{References for the literature data.}
\tablenotetext{\text{d}}{Value from the literature. Data observed using the VLA before the upgrade.}

\tablecomments{\footnotesize{The centimeter continuum information of sources G45.47$+$0.05 (at 6 cm) and Cepheus A were taken from the literature and the references are given in column 9. Also, for the analysis of IRAS 20126$+$4104, we use the radio images presented in \citet{2016ApJS..227...25R}. Units of right ascension are hours, minutes, and seconds, and units of declination are degrees, arcminutes, and arcseconds.
\\
     \osref{rxn:Urquhart2009}  \citet{2009A\string&A...501..539U};  \osref{rxn:Rosero2016}  \citet{2016ApJS..227...25R}; \osref{rxn:Curiel2006}  \citet{2006ApJ...638..878C} }}
\end{deluxetable}

\begin{deluxetable}{l c c c}
\tabletypesize{\scriptsize}
\tablecaption{VLA Calibrators  \label{VLA_cal}}
\tablewidth{0pt}
\tablehead{
\colhead{Calibrator}                  & 
\colhead{Astrometry Precision\tablenotemark{a}}  &
\colhead{Source Calibrated}    &
\colhead{Band}    \\[-20pt]\\}
\startdata
\setcounter{iso}{0}	
J0228$+$6721   & A  & AFGL 4029 & C, K, Q  \\
J0359$+$5057   & B  & AFGL 437  & C, K \\
J2230$+$6946   & A  & NGC 7538 IRS9 & C, K\\
J0735$-$1735   & A  & IRAS 07299$-$1651 & C\\
J1851$+$0035   & C  & G35.20$-$0.74 & C, K\\
J0228$+$6721   & A  & AFGL 437  & Q \\
J1924$+$1540   & A  & G45.47$+$0.05   & Q \\
J2250$+$5550   & A  & NGC 7538 IRS9 & Q\\
\enddata
\tablenotetext{\text{a}}{Astrometric precision A, B, and C correspond to positional accuracies of $<$2 mas, 2--10 mas, and 0.01--0.15 arcsecs, respectively.}

\end{deluxetable}

\subsection{VLA data}\label{images}

\subsubsection{6 cm Data}

The 6 cm (\emph{C-}band) observations were made in the A configuration
providing angular resolutions $\sim$$0\rlap.^{\prime \prime}$3 --
$0\rlap.^{\prime \prime}$5. The data for sources AFGL 4029, AFGL 437
and NGC 7538 IRS9 are from project code 12B--140 from observations
taken in 2012 and the data for source G35.20$-$0.74 are from project
code 13B--210 from observations taken in 2013. The data consist of
two 1 GHz wide basebands (8 bit samplers) centered at 5.3 and 6.3 GHz,
where each baseband was divided into 8 spectral windows (SPWs), each
with a bandwidth of 128~MHz. The data were recorded in 16 unique SPWs,
each comprised of 64 channels and each channel being 2 MHz wide,
resulting in a total bandwidth of 2048 MHz (before ``flagging''). 3C48
was used as flux density and bandpass calibrator for regions AFGL
4029 and AFGL 437, and 3C286 was used as flux density and bandpass
calibrator for regions G35.20$-$0.74 and NGC 7538 IRS9. For sources
AFGL 4029, AFGL 437 and NGC 7538 IRS9 the observations were made
alternating on a target source for $\sim$13 minutes and a phase
calibrator for $\sim$1 minute, for a total on-source time of $\sim $26
minutes. For G35.20$-$0.74 the observations were made alternating on a
target source for $\sim$10 minutes and a phase calibrator for $\sim$1
minute, for a total on-source time of $\sim $40 minutes.

The observations for source IRAS 07299$-$1651 are from project code
18A--294 (PI: Rosero) taken in 2018 and the data consists of two $\sim$2 GHz wide
basebands (3 bit samplers) centered at 5.03 and 6.98 GHz. The data
were recorded in 30 unique SPWs, each comprised of 64 channels and
each channel being 2 MHz wide, resulting in a total bandwidth of 3842
MHz (before ``flagging''). 3C48 was used as flux density and bandpass
calibrator and the observations were made alternating on a target
source for $\sim$9.5 minutes and a phase calibrator for $\sim$40
seconds, for a total on-source time of $\sim $37 minutes.

The data were processed using NRAO's Common Astronomy Software
Applications (CASA)\footnote{http://casa.nrao.edu} package. Eight
channels at the edges of each baseband were flagged due to substantial
band roll-off (and therefore loss of sensitivity). In addition, we
inspected the data for radio frequency interference (RFI) or other
problems, performing ``flagging'' when needed. The flux density scale
was set via standard NRAO models using the task {\tt setjy} for the
flux calibrators and using the \citet{2013ApJS..204...19P} flux
scale. We used the {\tt gencal} task to check for antenna position
corrections and also to apply gain curve and antenna efficiency
factors.  Delay and bandpass solutions were formed based on
observations of the flux density calibrator. These solutions were
applied when solving for the final amplitude and phase calibration
using the task {\tt gaincal} over the full bandwidth.  We measured the
flux density of the phase calibrators using the task {\tt
  fluxscale}. The amplitude, phase, delay and bandpass solutions were
applied to the target sources using the task {\tt applycal}. The
images were made using the \texttt{tclean} task and Briggs
\texttt{Robust}$= 0.5$ weighting. For source G35.20$-$0.74 we
performed self-calibration.
 
For regions AFGL 4029, AFGL 437, G35.20$-$0.74 and NGC 7538 IRS9, we
made two images, each of $\sim$1 GHz baseband composed of 8 SPWs and
also a combined image using data from both basebands with a total of
16 SPWs. For region IRAS 07299$-$1651 we made two images, each of
$\sim$2 GHz baseband composed of 15 SPWs and also a combined image
using data from both basebands with a total of 30 SPWs. All maps were
primary-beam-corrected. Table \ref{Observations_table} columns 5 and 6
show the synthesized beam (size and position angle) and the rms of
the combined images.

\subsubsection{1.3 cm Data}

The 1.3 cm (\emph{K-}band) observations were made in the B
configuration providing angular resolutions $0\rlap.^{\prime
  \prime}$3. The data reduced in this work is for source G35.20$-$0.74
(project code 13B--033) from observations taken in 2013.  The data
consists of two 4 GHz wide basebands (3 bit samplers) centered at 19.9
and 23.9 GHz, where each baseband was divided into 32 spectral windows
(SPWs), each with a bandwidth of 128 MHz. The data were recorded in 64
unique SPWs, each comprised of 128 channels and each channel being 1
MHz wide, resulting in a total bandwidth of 8192 MHz (before
``flagging''). 3C286 was used as the flux density and bandpass calibrator
for region G35.20$-$0.74. The observations were made alternating on a
target source for $\sim$2.5 minutes and a phase calibrator for $\sim$1
minute, for a total on-source time of $\sim $7 minutes.

The data reduction was done in the same fashion as that for the 6 cm
observations. In addition, we corrected for atmospheric opacity using
the weather station information from the \texttt{plotWeather} task and
creating the calibration table using {\tt gencal}. The images were
made using the \texttt{tclean} task and Briggs \texttt{Robust}$= 0.5$
weighting. We made two images, each of $\sim$ 4 GHz baseband composed
of 32 SPWs and also a combined image using data from both basebands
with a total of 64 SPWs. All maps were primary-beam-corrected. Table
\ref{Observations_table} columns 5 and 6 show the synthesized beam
(size and position angle) and the rms of the combined images.

\subsubsection{0.7 cm Data}

The 0.7 cm (\emph{Q-}band) observations were made in the A
configuration --- except for G35.20$-$0.74 where B configuration was
used --- providing angular resolutions $\sim$ $0\rlap.^{\prime
  \prime}$04 -- $0\rlap.^{\prime \prime}$06 and $\sim$
$0\rlap.^{\prime \prime}$2 for the B configuration data. The data for
sources AFGL 4029 and AFGL 437 are from project code 15A--238 from
observations taken in 2015, the data for source G35.20$-$0.74 is from
project code 13B--210 from observations taken in 2013, the data for
source G45.47$+$0.05 is from project code 14A--113 from observations
taken in 2014, and the data for source NGC 7538 IRS9 is from project
code 14A--092 from observations taken in 2014.  The data for sources
AFGL 4029, AFGL 437 and NGC 7538 IRS9 consist of two 4 GHz wide
basebands (3 bit samplers) centered at 41.9 and 45.9 GHz, where each
baseband was divided into 32 spectral windows (SPWs), each with a
bandwidth of 128 MHz. The data were recorded in 64 unique SPWs, each
comprised of 64 channels and each channel being 2 MHz wide, resulting
in a total bandwidth of 8192 MHz (before ``flagging''). The correlator
setup for the observations of sources G35.20$-$0.74 and G45.47$+$0.05
was the same as that used at 1.3 cm, with the two basebands centered
at 41.9 and 45.9 GHz. 3C48 was used as flux density and bandpass
calibrators for regions AFGL 4029, AFGL 437 and NGC 7538 IRS9, and
3C286 were used as flux density and bandpass calibrators for regions
G35.20$-$0.74 and G45.47$+$0.05. For sources AFGL 4029 and AFGL 437
the observations were made alternating on a target source for
$\sim$1.7 minutes and a phase calibrator for $\sim$1 minute, for a
total on-source time of $\sim $7 minutes. For G35.20$-$0.74 the
observations were made alternating on a target source for $\sim$2.5
minutes and a phase calibrator for $\sim$1 minute, for a total
on-source time of $\sim $38 minutes. For G45.47$+$0.05 the
observations were made alternating on a target source for $\sim$2
minutes and a phase calibrator for $\sim$1 minute, for a total
on-source time of $\sim $23 minutes. For NGC 7538 IRS9 the
observations were made alternating on a target source for $\sim$2
minutes and a phase calibrator for $\sim$1 minute, for a total
on-source time of $\sim $11 minutes.

The data reduction was done in the same fashion as that for the 6 cm
observations. In addition, we corrected for atmospheric opacity using
the weather station information from the \texttt{plotWeather} task and
creating the calibration table using {\tt gencal}. The images were
made using the \texttt{tclean} task and Briggs \texttt{Robust}$= 0.5$
weighting. For source G45.47$+$0.054 we performed self-calibration. We
made two images, each of $\sim$ 4 GHz baseband composed of 32 SPWs and
also a combined image using data from both basebands with a total of
64 SPWs. All maps were primary-beam-corrected. Table
\ref{Observations_table} columns 5 and 6 show the synthesized beam
(size and position angle) and the rms of the combined images.

\section{Theoretical models}\label{Tanaka_models}

\citetalias{2017ApJ...843...33D} investigated the protostellar
properties of the eight SOMA sources presented in this paper, fitting
the given set of infrared observations to the ZT model grid.  However,
the obtained properties of the IR-only SED models still have
relatively large allowed variations in their parameters, i.e.,
significant degeneracies.  In this paper, we aim to constrain the
models further by comparing the radio observations and the
\citetalias{2016ApJ...818...52T} model of free-free emission.  In this
section, we briefly revisit these theoretical models, i.e., the ZT and
\citetalias{2016ApJ...818...52T} models, and the model selection
methods.

\subsection{ZT model grid for dust thermal emission}

In high-mass star formation, the vast majority of the energy emitted
from the protostar and the innermost accretion flow is at optical and
UV wavelengths.  However, it is hard to directly detect the radiation
from the protostar and its vicinity because most of the flux emitted at
those wavelengths is immediately absorbed by the surrounding material
and reemitted in the infrared.  Therefore, in order to investigate the
properties of the embedded protostar, theoretical synthetic
observational modeling is necessary.  In a series of papers
\citep{2011ApJ...733...55Z,2013ApJ...766...86Z,2014ApJ...788..166Z,2018ApJ...853...18Z}
a theoretical model of the evolution of high-mass protostars based on
the Turbulent Core scenario \citep{2003ApJ...585..850M} has been
developed, which provides the continuum flux at infrared and optical
wavelengths.  These are referred to as the ZT models.

In the ZT models, massive protostars are assumed to form from
massive prestellar cores.  The prestellar core properties are
parameterized by the initial core mass $M_c$ and the mass surface
density of the ambient clump $\Sigma_{\rm cl}$.  The latter determines
the surface pressure of cores, and thus sets their sizes and densities
together with $M_c$.  The core undergoes collapse that forms a
protostar-disk system at its center.  The infalling structure is given
by the self-similar solution \citep{1997ApJ...476..750M} including
effects of rotation \citep{1976ApJ...210..377U}.  The disk structure
is described with an $\alpha$-disk solution
\citep{1973A&A....24..337S}, with an improved treatment including the
infall to and outflow from the disk.  The accretion rate onto the
protostar is regulated by the feedback of the MHD disk wind
\citep{1982MNRAS.199..883B,2000ApJ...545..364M}.  The protostellar
evolution is calculated consistently to the accretion rate based on
the method developed by \citet{2009ApJ...703.1810H} and
\citet{2010ApJ...721..478H}, solving the basic stellar structure
equations, i.e., continuity, hydrostatic balance, energy conservation
and transfer.

The evolution of the protostar and its surrounding gas structure are
calculated self-consistently from the initial core parameters, i.e.,
$M_c$ and $\Sigma_{\rm cl}$.  The protostellar mass $m_*$ is used as
the third parameter to specify a particular stage of the evolutionary
tracks.  The continuum radiative transfer calculations at infrared and
optical wavelengths have been performed for $432$ physical models
defined by different sets of $\left( M_c,\:\Sigma_{\rm cl},\:m_*
\right)$ using the latest version of the Monte Carlo code, HOCHUNK3d
\citep{2003ApJ...591.1049W,2013ApJS..207...30W}.  For each model, 20
inclination angles $\theta_{\rm view}$ are sampled evenly in cosine
space to produce the SEDs.  To compare with the observations, a
variable foreground extinction $A_{V}$ is applied to the model SEDs.
Thus, for a given source distance, a set of five parameters of $\left(
M_c,\:\Sigma_{\rm cl},\:m_*,\:\theta_{\rm view},\:A_{V}\right)$ gives
one SED in the ZT model grid.

In \citetalias{2017ApJ...843...33D}, the authors searched for the best
fit models for the eight SOMA regions presented in this paper from the
ZT model grid to investigate the properties of their protostars and
surrounding gas.  They used $\chi^2$ minimization (in log space) to
find the best models to fit a given set of observations at
$8$--$70\:\micron$ for the eight SOMA sources.  They successfully
found models that can explain the infrared observed SED of each of
these SOMA sources.  However, even among the best five models, the
obtained protostellar properties, such as stellar masses and
bolometric luminosities, have relatively large degeneracies.

\subsection{\citetalias{2016ApJ...818...52T} model for free-free emission}

To further constrain the protostellar properties we use the additional
diagnostic of free-free emission from photoionized gas.  The UV flux
dramatically increases during the Kelvin-Helmholtz (KH) contraction
phase in the formation of a high-mass star, although this cannot be
directly observed due to absorption by the surrounding gas and dust.
However, the gas absorbing Lyman continuum photons with $>13.6\rm\:eV$
becomes ionized and emits thermal bremsstrahlung, i.e., free-free
emission.  Therefore, free-free emission at radio wavelengths from the
photoionized region can provide the direct signpost of the
evolutionary state of the embedded protostar.

Using the physical parameters of the best five models for the
infrared, we calculated the photoionized structures and the free-free
flux from them using the method of \citetalias{2016ApJ...818...52T}.
The basic protostellar properties and the density structure of the
surrounding material are given from the ZT model grids.  The ionizing
photon rate is evaluated using the stellar atmosphere model of
\citet{2004astro.ph..5087C}.  The temperature of the ionized gas is
evaluated by C{\small LOUDY} \citep{2013RMxAA..49..137F}.  The
photoionized structure is obtained by the ray-tracing radiative
transfer calculation \citep{1992ApJS...80..819S,2013ApJ...773..155T},
which allows a treatment of both the direct and diffuse ionizing
radiation fields.

Some hydrodynamical simulations have included the MHD outflow feedback in high-mass star formation \citep[e.g.,][]{2016ApJ...832...40K, 2017MNRAS.470.1026M}. Especially, \citet{2018A&A...616A.101K} recently performed the first simulations self-consistently including the protostellar outflow together with the radiation force and the photoionization feedback. However, since they focused on the dynamical impact of feedback processes, they have not conducted the observational modeling for dust and free-free emissions.  Moreover, they explored only two initial conditions, while our semi-analytic models span a wide range of environmental conditions (\citetalias{2016ApJ...818...52T}; \citealt{2017ApJ...849..133T}).  We note that the predictions of our semi-analytic models, such as the star formation efficiencies, are quantitatively consistent with those from simulations by \citet{2018A&A...616A.101K}, which supports the accuracy of our semi-analytic models.


\begin{deluxetable}{l c c c c c c c }

\tabletypesize{\scriptsize}
\tablecaption{AFGL 4029: Parameters from Radio Continuum   \label{AFGL4029_par}}
\tablewidth{0pt}
\tablehead{
\colhead{Scale} &
\colhead{R.A.}   &
\colhead{Dec. }               & 
\colhead{S$_{5.3\,\,\text{GHz}}$}  & 
\colhead{S$_{6.3\,\,\text{GHz}}$}  & 
\colhead{S$_{41.9\,\,\text{GHz}}$}  & 
\colhead{S$_{45.9\,\,\text{GHz}}$} &
\colhead{Spectral}            
 \\[2pt]
\colhead{}                            & 
\colhead{(J2000)}                      & 
\colhead{(J2000)}                      & 
\colhead{(mJy)}                      & 
\colhead{(mJy)}                      & 
\colhead{(mJy)}                      & 
\colhead{(mJy)}                       &      
\colhead{Index}                          \\[-20pt]\\}
\startdata
\setcounter{iso}{0}	
SOMA            & 03:01:31.28       & $+$60.29.12.9     & 0.44(0.52) & 0.42(0.77) & 4.73(30.04) & $<$134.24  & $<$1.1        \\[3pt] 
Intermediate    & 03:01:31.28       & $+$60.29.12.9     & 0.31(0.06) & 0.34(0.08) & 0.95(3.12) & $<$14.10  & $<$0.5        \\[3pt] 
Inner           & 03:01:31.28       & $+$60:29:12.8     & 0.14(0.02) & 0.16(0.03) & 0.74(0.11) & 0.48(0.11) & 0.7(0.1)       \\[3pt] 

\enddata

\tablecomments{\footnotesize{The intermediate scale corresponds to the extend of the radio jet.  Units of right ascension are hours, minutes, and seconds, and units of declination are degrees, arcminutes, and arcseconds.
 }}
\end{deluxetable}


\begin{deluxetable}{l c c c c c c c }
\tabletypesize{\scriptsize}
\tablecaption{AFGL 437: Parameters from Radio Continuum   \label{AFGL437_par}}
\tablewidth{0pt}
\tablehead{
\colhead{Scale} &
\colhead{R.A.}   &
\colhead{Dec. }               & 
\colhead{S$_{5.3\,\,\text{GHz}}$}  & 
\colhead{S$_{6.3\,\,\text{GHz}}$}  & 
\colhead{S$_{41.9\,\,\text{GHz}}$}  & 
\colhead{S$_{45.9\,\,\text{GHz}}$} &
\colhead{Spectral}           
 \\[2pt]
\colhead{}                            & 
\colhead{(J2000)}                      & 
\colhead{(J2000)}                      & 
\colhead{(mJy)}                      & 
\colhead{(mJy)}                      & 
\colhead{(mJy)}                      & 
\colhead{(mJy)}                       &
\colhead{Index}                               \\[-20pt]\\}
\startdata
\setcounter{iso}{0}	
SOMA            & 03:07:24.49       & $+$58.30.42.8     & 0.82(0.28) & 0.36(0.57) & 0.28(10.34) & 3.37(15.54) & -4.7(3.6)       \\[3pt] 
Intermediate    & \nodata           & \nodata           & \nodata & \nodata & \nodata & \nodata & \nodata       \\[3pt] 
Inner           & 03:07:24.49       & $+$58:30:42.8     & 0.77(0.09) & 0.80(0.09) & 1.39(0.20) & 1.57(0.30) & 0.3(0.1)       \\[3pt] 

\enddata

\tablecomments{\footnotesize{ Units of right ascension are hours, minutes, and seconds, and units of declination are degrees, arcminutes, and arcseconds.
 }}
\end{deluxetable}

\vspace{-.2cm}
\begin{deluxetable}{l c c c  c c}
\tabletypesize{\scriptsize}
\tablecaption{IRAS 07299$-$1651: Parameters from Radio Continuum }\label{I07_par}
\tablewidth{0pt}
\tablehead{
\colhead{Scale} &
\colhead{R.A.}   &
\colhead{Dec. }               & 
\colhead{S$_{5.0\,\,\text{GHz}}$}  &
\colhead{S$_{7.0\,\,\text{GHz}}$}  &
\colhead{Spectral}         
 \\[2pt]
\colhead{}                            & 
\colhead{(J2000)}                      & 
\colhead{(J2000)}                      & 
\colhead{(mJy)}                          &
\colhead{(mJy)}                          &
\colhead{Index}                           \\[-20pt]\\}
\startdata
\setcounter{iso}{0}	
SOMA            & 07:32:09.74       & -16.58.11.3       & 1.56(0.25) & 1.62(0.34) & 0.1(0.8)       \\[3pt] 
Intermediate    & \nodata           & \nodata           & \nodata & \nodata & \nodata       \\[3pt] 
Inner           & 07:32:09.79       & $-$16:58:10.9     & 1.15(0.12) & 1.47(0.16) & 0.7(0.5)       \\[3pt] 
\enddata
\tablecomments{\footnotesize{Units of right ascension are hours, minutes, and seconds, and units of declination are degrees, arcminutes, and arcseconds.}}
\end{deluxetable}


\begin{deluxetable}{l c c c c c c c c c}
\tabletypesize{\scriptsize}
\tablecaption{G35.20-0.74: Parameters from Radio Continuum   }\label{G35_par}
\tablewidth{0pt}
\tablehead{
\colhead{Scale} &
\colhead{R.A.}   &
\colhead{Dec. }               & 
\colhead{S$_{4.9\,\,\text{GHz}}$}  & 
\colhead{S$_{6.9\,\,\text{GHz}}$}  & 
\colhead{S$_{19.9\,\,\text{GHz}}$}  & 
\colhead{S$_{23.9\,\,\text{GHz}}$}  & 
\colhead{S$_{41.9\,\,\text{GHz}}$}  & 
\colhead{S$_{45.9\,\,\text{GHz}}$}   &
\colhead{Spectral}        
 \\[2pt]
\colhead{}                            & 
\colhead{(J2000)}                      & 
\colhead{(J2000)}                      & 
\colhead{(mJy)}                      & 
\colhead{(mJy)}                      & 
\colhead{(mJy)}                      & 
\colhead{(mJy)}                      & 
\colhead{(mJy)}                      & 
\colhead{(mJy)}                       &
\colhead{Index}                               \\[-20pt]\\}
\startdata
\setcounter{iso}{0}	
SOMA            & 18:58:13.02       & $+$01.40.36.2     & 15.08(1.77) & 13.51(1.77) & 11.73(5.32) & 13.43(8.85) & 10.80(13.97) & 5.68(25.52) & -0.2(0.3)       \\[3pt] 
Intermediate    & 18:58:13.02       & $+$01.40.36.2     & 14.46(1.45) & 12.88(1.30) & 14.53(1.52) & 15.15(1.65) & 7.05(1.09) & 6.22(1.49) & -0.2(0.1)       \\[3pt] 
Inner           & 18:58:13.04       & $+$01:40:35.9     & 0.74(0.12) & 0.79(0.10) & 1.82(0.21) & 2.27(0.26) & 2.52(0.26) & 3.08(0.32) & 0.7(0.1)       \\[3pt] 

\enddata

\tablecomments{\footnotesize{The intermediate scale corresponds to the extend of the radio jet.  Units of right ascension are hours, minutes, and seconds, and units of declination are degrees, arcminutes, and arcseconds. 
 }}
\end{deluxetable}



\begin{deluxetable}{l c c c c c c}
\tabletypesize{\scriptsize}
\tablecaption{G45.47$+$0.05: Parameters from Radio Continuum   \label{G45_par}}
\tablewidth{0pt}
\tablehead{
\colhead{Scale} &
\colhead{R.A.}   &
\colhead{Dec. }               & 
\colhead{S$_{5\,\,\text{GHz}}$}  &
\colhead{S$_{41.9\,\,\text{GHz}}$}  &
\colhead{S$_{45.9\,\,\text{GHz}}$}  &
\colhead{Spectral}        
 \\[2pt]
\colhead{}                            & 
\colhead{(J2000)}                      & 
\colhead{(J2000)}                      & 
\colhead{(mJy)}                      & 
\colhead{(mJy)}                       &
\colhead{(mJy)}                       &
\colhead{Index}                               \\[-20pt]\\}
\startdata
\setcounter{iso}{0}	
SOMA            & 19:14:25.67       & $+$11.09.25.4     & 58.00(13.52) & 140.10(45.80) & 157.21(78.91) & 0.4(0.2)       \\[3pt] 

Intermediate    & \nodata           & \nodata           & \nodata      & \nodata       & \nodata       & \nodata        \\[3pt] 

Inner           & 19:14:25.67       & $+$11:09:25.9     & 91.00(9.75)  & 102.90(11.27) & 119.70(13.05) & 0.1(0.1)       \\[3pt] 

\enddata

\tablecomments{\footnotesize{The quoted flux at 5 GHz is from  \citet{2009A\string&A...501..539U}. The quoted fluxes at the inner scale for the higher frequencies correspond to the southern source. Units of right ascension are hours, minutes, and seconds, and units of declination are degrees, arcminutes, and arcseconds.
 }}
\end{deluxetable}


\begin{deluxetable}{l c c c c c c c c}
\tabletypesize{\scriptsize}
\tablecaption{IRAS 20126$+$4104: Parameters from Radio Continuum   \label{I20_par}}
\tablewidth{0pt}
\tablehead{
\colhead{Scale} &
\colhead{R.A.}   &
\colhead{Dec. }               & 
\colhead{S$_{4.9\,\,\text{GHz}}$}  & 
\colhead{S$_{7.4\,\,\text{GHz}}$}  & 
\colhead{S$_{20.9\,\,\text{GHz}}$}  & 
\colhead{S$_{25.5\,\,\text{GHz}}$}   &
\colhead{Spectral}         
 \\[2pt]
\colhead{}                            & 
\colhead{(J2000)}                      & 
\colhead{(J2000)}                      & 
\colhead{(mJy)}                      & 
\colhead{(mJy)}                      & 
\colhead{(mJy)}                      & 
\colhead{(mJy)}                       &
\colhead{Index}                               \\[-20pt]\\}
\startdata
SOMA            & 20:14:26.05       & $+$41.13.32.5     & 0.37(0.54) & 0.50(0.75) & 0.29(1.49) & 1.13(2.76) & 0.2(1.9)       \\[3pt] 
Intermediate    & \nodata           & \nodata           & \nodata & \nodata & \nodata & \nodata & \nodata       \\[3pt] 
Inner           & 20:14:26.03       & $+$41:13:32.5     & 0.06(0.02) & 0.08(0.02) & 0.64(0.07) & 0.85(0.09) & 1.8(0.1)       \\[3pt] 

\enddata
\tablecomments{\footnotesize{Units of right ascension are hours, minutes, and seconds, and units of declination are degrees, arcminutes, and arcseconds.
}}
\end{deluxetable}

\clearpage
\begin{deluxetable}{l c c c c c c c c c }
\tabletypesize{\scriptsize}
\tablecaption{Cepheus A: Parameters from Radio Continuum   \label{CepA_par}}
\tablewidth{0pt}
\tablehead{
\colhead{Scale} &
\colhead{R.A.}   &
\colhead{Dec. }               & 
\colhead{S$_{1.5\,\,\text{GHz}}$}  &
\colhead{S$_{4.9\,\,\text{GHz}}$}  &
\colhead{S$_{8.3\,\,\text{GHz}}$}  &
\colhead{S$_{14.9\,\,\text{GHz}}$}  & 
\colhead{S$_{23.1\,\,\text{GHz}}$}  & 
\colhead{S$_{43.0\,\,\text{GHz}}$}   &
\colhead{Spectral}         
 \\[2pt]
\colhead{}                            & 
\colhead{(J2000)}                      & 
\colhead{(J2000)}                      & 
\colhead{(mJy)}                      & 
\colhead{(mJy)}                      &
\colhead{(mJy)}                      &
\colhead{(mJy)}                      &
\colhead{(mJy)}                      & 
\colhead{(mJy)}                       &
\colhead{Index}                               \\[-20pt]\\}
\startdata
\setcounter{iso}{0}	
SOMA            & \nodata           & \nodata           & \nodata  & \nodata   & \nodata   & \nodata   & \nodata   & \nodata   & \nodata  \\[3pt] 
Intermediate    & 22:56:17.98       & $+$62.01.49.4     & 3.40(0.10) & 7.50(0.10)  & 9.80(0.10)  & 15.80(0.20) & \nodata   & 35.00(2.00) & 0.66(0.01)       \\[3pt] 
Inner           & 22:56:17.99       & $+$62:01:49.6     & \nodata  & \nodata   & 6.85(0.07)  & \nodata   & 18.50(0.30) & 65.00(0.50) & 1.38(0.01)       \\[3pt] 

\enddata
\tablecomments{\footnotesize{The parameters shown in the table are from \citet{1994ApJ...430L..65R} and \citet{2006ApJ...638..878C} for the intermediate and the inner scale, respectively. Units of right ascension are hours, minutes, and seconds, and units of declination are degrees, arcminutes, and arcseconds.
 }}
\end{deluxetable}

\begin{deluxetable}{l c c c c c c  c}
\tabletypesize{\scriptsize}
\tablecaption{NGC 7538 IRS9: Parameters from Radio Continuum   \label{NGC_par}}
\tablewidth{0pt}
\tablehead{
\colhead{Scale} &
\colhead{R.A.}   &
\colhead{Dec. }               & 
\colhead{S$_{5.3\,\,\text{GHz}}$}  & 
\colhead{S$_{6.3\,\,\text{GHz}}$}  & 
\colhead{S$_{41.9\,\,\text{GHz}}$}  & 
\colhead{S$_{45.9\,\,\text{GHz}}$}  &
\colhead{Spectral}          
 \\[2pt]
\colhead{}                            & 
\colhead{(J2000)}                      & 
\colhead{(J2000)}                      & 
\colhead{(mJy)}                      & 
\colhead{(mJy)}                      & 
\colhead{(mJy)}                      & 
\colhead{(mJy)}                      &
\colhead{Index}                                \\[-20pt]\\}
\startdata
\setcounter{iso}{0}	
SOMA            & 23:14:01.77       & $+$61.27.19.8     & 5.00(1.68) & 25.74(5.46) & 2.47(103.89) & $<$515.49  & $<$1.1        \\[3pt] 
Intermediate    & \nodata           & \nodata           & \nodata & \nodata & \nodata & \nodata & \nodata       \\[3pt] 
Inner           & 23:14:01.76       & $+$61:27:19.8     & 0.42(0.04) & 0.48(0.06) & 2.50(0.29) & 2.47(0.29) & 0.8(0.1)       \\[3pt] 

\enddata

\tablecomments{\footnotesize{Units of right ascension are hours, minutes, and seconds, and units of declination are degrees, arcminutes, and arcseconds.
 }}
\end{deluxetable}

\section{Results}\label{results}

In this paper, a radio detection is defined when the peak intensity
$I_{\nu}$ is $\geq$5 times the image rms ($\sigma$) in either of the
baseband-combined images (see \S\ref{images}) at the different bands
(i.e., \textit{C-, K-} or \textit{Q-}band). For non-detections in one
of the combined images we report a 3$\sigma$ limit value for the flux
density at the given frequency. Figure \ref{fig:VLA_contours} shows
VLA contour plots of the C-band: 6 cm (red), K-band: 1.3 cm (magenta) and Q-band: 0.7 cm
(blue) combined images toward all the radio sources detected in our
sample and overlaid to {\it SOFIA}-FORCAST 37$\mu$m images.  The blue
circles represent the SOMA apertures set from 70 $\mu$m images (except
for region IRAS$\,$07299$-$1651 where the aperture radius is set from
the 37.1 $\mu$m image) reported by \citetalias{2017ApJ...843...33D}
and used to build their IR SEDs (see \citetalias{2017ApJ...843...33D}
Table~2).  The astrometric accuracy of the infrared images presented
in \citetalias{2017ApJ...843...33D} (including {\it SOFIA}, \emph{Herschel}
and \emph{Spitzer} data) and the VLA data presented here are better
than $0\rlap.^{\prime \prime}$5 and $0\rlap.^{\prime \prime}$1 (see
Table \ref{VLA_cal}), respectively.

%
%

\begin{deluxetable}{l c c  }
\tabletypesize{\scriptsize}
\tablecaption{SOMA and Intermediate Scales  \label{scales}}
\tablewidth{0pt}
\tablehead{
\colhead{Region}                  & 
\colhead{SOMA} & 
\colhead{Intermediate}    \\
\colhead{} &
\colhead{R ($^{\prime \prime}$)}&
\colhead{w($^{\prime \prime}$) $\times$ h($^{\prime \prime}$)}\\[-20pt]\\}
\startdata
\setcounter{iso}{0}	
AFGL 4029                & 11.2 & 3.37 $\times$ 1.27   \\
AFGL 437                  & 3.84 & \nodata \\
IRAS 07299$-$1651  & 7.7  & \nodata \\
G35.20$-$0.74           & 32.0  & 3.36 $\times$ 16.15\\
G45.47$+$0.05          &14.4  & \nodata \\
IRAS 20126$+$4104  &12.8  & \nodata \\
Cepheus A                  & 48.0  & \nodata\tablenotemark{a} \\
NGC 7538 IRS9         & 25.6  & \nodata \\
\enddata
\tablenotetext{\text{a}}{Size changes with frequency and they are reported in \citet{1994ApJ...430L..65R} Table 1.}
\tablecomments{\footnotesize{The reported values correspond to a circle of radius R for the SOMA scale and a box of height h and width w for the intermediate scale.
 }}

\end{deluxetable}

Tables \ref{AFGL4029_par} to \ref{NGC_par} report the radio parameters
for each of the eight regions studied in this paper. These parameters
were measured based on different size scales as follows. The
\emph{SOMA} scale refers to the size of the aperture radius used by
\citetalias{2017ApJ...843...33D} to measure their IR fluxes (except
for source AFGL 437; see below).  The \emph{Intermediate} scale is
based on the morphology of the radio source, specifically if the
detections appear of jet-like nature.  The \emph{Inner}
scale is given
by the size of the central radio detection that is likely most closely
associated with the driving protostar (i.e., association with compact
millimeter dust continuum emission).
For the presented VLA data we determined the flux density
$S_{\nu}$ in each wide band image in the \emph{SOMA} scale and in the
\emph{Intermediate} scale by using the task \texttt{imstat} of CASA
either enclosing the SOMA aperture in a circular region or enclosing
the elongated jet-like structure in a box, respectively (see Table \ref{scales}). The
uncertainties of the flux density for these two scales are estimated
as $\sigma_{image} (npts/beam\,area)^{0.5}$ added in quadrature with
an assumed 10$\%$ error in calibration, where $\sigma_{image}$ is the
rms of the image, \emph{npts} is the number of pixels enclosed in the
box or the circular region and the \emph{beam area} is the number of
pixels within a synthesized beam of the image. For the \emph{Inner}
scale we determined the flux density using the task \texttt{imfit} of
CASA, and the uncertainty was estimated using the statistical error
from the Gaussian fit added in quadrature with an assumed 10$\%$ error
in calibration.

Tables \ref{AFGL4029_par} to \ref{NGC_par} column 1 show the given
scale and for each scale columns 2 and 3 report the R.A. and decl.,
which for the \emph{SOMA} scale they refer to the pointing center
observations of {\it SOFIA}-FORECAST (see \citetalias{2017ApJ...843...33D}
Table~1), for the \emph{Intermediate} scale they refer to a middle
point in the jet-like detection and for the \emph{Inner} scale they
refer to the R.A. and decl. of the peak intensity of the central
detected object. The following columns are the flux densities
($S_{\nu}$) at different frequencies with the uncertainties given in
parentheses, and the last columns in Tables \ref{AFGL4029_par} to
\ref{NGC_par} report the spectral indices and their uncertainties at
each scale (see \S \ref{spectral_indices}). Since the radio data are not sensitive to extended emission over scales as large as the \emph{SOMA} and possibly the \emph{Intermediate} scales, the flux measurements represent the sum over all compact sources within the scale and the spectral indices contain some corresponding uncertainty. The error bars for these measurements are large  due to having many independent beams within the scale.

\subsection{Morphology and Multiplicity}

All the target regions presented in this paper have been detected in
cm continuum and we describe their morphology as either compact if the
detection shows no structure on the scale of a few synthesized beams
or extended otherwise. Below we describe the centimeter wavelength
detections toward each target; for a detailed background on each of
these regions see \citetalias{2017ApJ...843...33D}.

\subsubsection{AFGL 4029}

AFGL 4029 is composed of two mid-IR sources, IRS1 and IRS2, with the
former being the source of interest in this work. In our data we
detected at least four centimeter continuum sources at C-band and one at
Q-band, being the eastern and central sources with 5$\sigma$ detections
(also reported by \citealt{2001RMxAA..37...83Z}) and the two western
ones being 3$\sigma$ detections. We detected only one of the sources of
the binary system found by \citet{2001RMxAA..37...83Z} at 3.6 cm
(named by them as G138.295$+$1.555 S and located at
RA(J2000)=$03^{h}01^{m}31{\rlap.}^{s}273$,
Dec(J2000)$=+$$60^{\circ}29^{\prime}12{\rlap.}^{\prime\prime}80 $)
even though our observations have $\sim$2 times higher sensitivity and
a similar resolution ($\sim$ $0\rlap.^{\prime \prime}$3) as
theirs. This may be further indication that the source
G138.295$+$1.555 N (not detected in our analyzed data), which they
reported to be at a separation of $\sim$ $0\rlap.^{\prime \prime}$6
(or 1200 au at the distance of the region) from G138.295$+$1.555 S, is
a variable radio source, as also suggested by \citet{2001RMxAA..37...83Z}.

Moreover, we detected the extended east-west fainter emission that
they interpreted as being part of an ionized jet that is emanating
from G138.295$+$1.555 S (our detected radio source at the Inner scale)
based on the morphology and their alignment with larger scale
outflows (e.g., optical jet: \citealt{1990ApJ...357L..45R} and CO
molecular outflow:
\citealt{2011MNRAS.418.2121G}). \citet{2001RMxAA..37...83Z} reported a
flux density of $\sim$0.25 mJy at 3.6 cm for G138.295$+$1.555 S and
detected it as an unresolved source at a resolution of $\sim$
$0\rlap.^{\prime \prime}$3. The presented data provide further evidence
that this system corresponds to an ionized jet (with a projected length
of $\sim$5000 au) based on the morphology, weak cm continuum emission
and a spectral index $\alpha \sim$0.7, which is consistent with the
typical spectral index of ionized jets (e.g.,
\citealt{1986ApJ...304..713R,1998AJ....116.2953A};
\citetalias{2016ApJ...818...52T}).  Additionally, although the three
other components of the jet (eastern and western components) are not
detected at Q-band, probably due to a lack of sensitivity at that band,
our upper limit estimates of their spectral indices ($\alpha
\lesssim$1) are still consistent with the expected values for ionized
jets.

\subsubsection{AFGL 437}

AFGL 437 is an infrared star-forming region composed of at least four
IR sources \citep[e.g.,][]{1981ApJ...246..801W}, where sources AFGL
437W and AFGL 437S have been the only ones reported to be associated
with centimeter continuum emission.  In our study towards this region,
AFGL 437W and AFGL 437S are detected at $> 5 \sigma$ as extended and
compact radio sources, respectively. Also, for the first time (to the
knowledge of the authors) we  detected very compact and unresolved centimeter continuum
towards  the very
embedded IR source WK 34 which is associated with AFGL 437N. 
The position of its
peak intensity is 
RA(J2000)=$03^{h}07^{m}24{\rlap.}^{s}571$,
Dec(J2000)$=+$$58^{\circ}30^{\prime}53{\rlap.}^{\prime\prime}00$.
WK 34 is very weak at 5.8 GHz with a flux density of
$\sim$  60 $\mu$Jy and at 44 GHz it has a flux density of $\sim$  700 $\mu$Jy, giving a positive spectral index of $\sim$ 1.2.
\citet{1996ApJ...458..670W}  speculated that WK 34 is tracing an outflow
cavity and \citet{2010MNRAS.402.2583K} suggested that WK 34 is a high-mass protostar in an earlier stage than AFGL 437W and AFGL 437S. Therefore, our high-sensitivity observations provide further evidence that high-mass stars even at the earlier stages are associated with very weak thermal radio continuum emission likely associated with ionized jets. 

Moreover, a CO molecular
outflow roughly oriented N-S has been reported
\citep{1992ApJ...397..492G, 2008A&A...484..361Q} towards the center of
the IR-emitting region, but it has a very low degree of collimation,
perhaps due to the superposition of multiple unresolved outflows
\citep{2012MNRAS.419.3338M}.
AFGL 437W and AFGL 437S have been previously observed
at 3.6 cm and 2 cm using the VLA by \citet{1992ApJ...392..616T} and
\citet{2012MNRAS.419.3338M} with angular resolutions of $\sim$
4$^{\prime \prime}$ and $\sim$1$^{\prime \prime}$, respectively.  AFGL
437W is a resolved extended source with an estimated total flux
density of $\sim$ 18 mJy at 3.6 cm and $\sim$ 17 mJy at 2 cm and
$\sim$ 22 mJy at 3.6 cm and $\sim$ 28 mJy at 2 cm from
\citet{1992ApJ...392..616T} and \citet{2012MNRAS.419.3338M},
respectively.  AFGL 437W appears to be an optically thin UC HII region
likely ionized by an early B-type ZAMS star
\citep{1992ApJ...392..616T}, although \citet{2012MNRAS.419.3338M}
suggested that this source
is a radio jet based on their detected morphology at 2 cm.  Since AFGL 437W is resolved and extended, the
archival data at 6 cm is missing part of its flux, and the higher
resolution data at 0.7 cm are also resolving out this source.

\citet{1992ApJ...392..616T} and \citet{2012MNRAS.419.3338M} estimated
a total flux density for AFGL 437S of $\sim$ 1 mJy at 3.6 cm and 2 cm
and $\sim$ 1.5 mJy at 3.6 cm and $<$ 2 mJy at 2 cm, respectively.  In
our data, AFGL 437S appears to be slightly elongated in the NE-SW
direction at longer wavelengths and it has a rising spectral index,
consistent with \citet{2012MNRAS.419.3338M}, thus AFGL 437S could be
either an optically thick UC/HC HII region or an ionized jet. In this
study, we test the scenario that AFGL 437S is associated with an
ionized jet that is likely undergoing mass loss
\citep{2012MNRAS.419.3338M}. In order to do so, we have built the SED
for AFGL 437S and therefore we do not use the photometry measured by
\citetalias{2017ApJ...843...33D} whose SED includes emission from all
the four IR sources of the cluster. Our \emph{SOMA} scale for AFGL
437S has a circular aperture of radius $R_{\rm ap}=$ 3.84$^{\prime
  \prime}$ around AFGL 437S and we only use the data at wavelengths
$<$40$\mu$m because at longer wavelengths the source is
unresolved. Table~\ref{AFGL_photometry}
has the integrated flux densities for AFGL 437S from
the IR data and Table \ref{AFGL437_par} shows our measured total flux
density for the centimeter wavelengths archival data.

\begin{deluxetable}{l c c c }
\tabletypesize{\scriptsize}
\tablecaption{AFGL 437S: Infrared Flux Densities  \label{AFGL_photometry}}
\tablewidth{0pt}
\tablehead{
\colhead{Facility}                  & 
\colhead{$\lambda$}                  & 
\colhead{$F_{\lambda,b\_sub}$\tablenotemark{a}}  &
\colhead{$F_{\lambda}$\tablenotemark{a}}     \\[2pt]
\colhead{}                            & 
\colhead{($\mu$m)}                      & 
\colhead{(Jy)}                      & 
\colhead{(Jy)}                           \\[-20pt]\\}
\startdata
\setcounter{iso}{0}	
\emph{Spitzer}/IRAC   & 3.6 & 0.35 & 0.40 \\
\emph{Spitzer}/IRAC   & 4.5 & 0.71 & 0.78 \\
\emph{Spitzer}/IRAC   & 5.8 & 1.56 & 1.77 \\
SOFIA/FORCAST   & 7.7 & 3.13 & 3.89 \\
\emph{Spitzer}/IRAC   & 8.0 & 2.77 & 3.32 \\
SOFIA/FORCAST    & 19.7 & 22.37 & 27.09 \\
SOFIA/FORCAST    & 31.5 & 46.27 & 65.10 \\
SOFIA/FORCAST    & 37.1 & 53.79 & 76.69 \\
\enddata
\tablenotetext{\text{a}}{Flux density derived with a fixed aperture radius of  $R_{ap}=$ 3.84$^{\prime \prime}$ from the 70$\mu$m data.  }

\tablecomments{\footnotesize{$F_{\lambda,b\_sub}$ and $F_{\lambda}$ correspond to the flux density derived with and without background subtraction, respectively.}}
\end{deluxetable}

\subsubsection{IRAS 07299$-$1651}

IRAS 07299$-$1651 appears extended in the NIR and MIR, it is
associated with methanol maser emission that shows a velocity gradient and
evidence of CO outflow wings, and all these tracers share a similar
elongation axis in the NW-SE direction
\citep{1999MNRAS.309..905W,2001MNRAS.326...36W}.

We present 6 cm observations towards this region where we have
detected at least one resolved and slightly elongated source. Also, we
detect at least two more unresolved sources that appear aligned with
the central source in the E-W direction.
However, our resulting images towards this region have calibration
errors and the source is very faint for self-calibration, thus we can
not rule out that these two sources are image artifacts.  The central
detection is consistent with the source reported by
\citet{1998MNRAS.301..640W} within their absolute positional accuracy
($\sim$ 1$\rlap.^{\prime \prime}$0) and they reported that it is associated with  6.7
GHz methanol masers. At this point, the nature of
IRAS 07299$-$1651 is unclear: it could be either an UC/HC HII
region or an ionized jet. Thus we require additional information such
as high resolution centimeter continuum at higher frequencies to
further constrain its nature.

\subsubsection{G35.20$-$0.74}

Also known as G35.20$-$0.74N, this region is seen in the infrared as
an elongated source oriented N-S that hosts a weakly collimated
bipolar CO molecular outflow that extends in the NE-SW direction. This
region is associated with at least three hot molecular cores labeled A,
B and C which are oriented in the NW-SE direction
\citep{2014A&A...569A..11S}. Centimeter continuum observations
\citep[e.g.,][]{2003MNRAS.339.1011G} have revealed the presence of a
string of radio sources that are coincident with the N-S elongated IR
emission seen in the region. Recently, \citet{2016A&A...593A..49B}
detected 17 centimeter continuum sources at angular resolutions of $\sim$
0$\rlap.^{\prime \prime}$4 -- $\sim$ 0$\rlap.^{\prime \prime}$05 (some
of these data are also used in our study) towards G35.20$-$0.74 and
based on the spectral indices of each component found that core B is
associated with an UC/HC HII region (their source 8a) that is part of a
young binary system ($\sim$ 0$\rlap.^{\prime \prime}$37 apart) and it
is likely driving an ionized jet oriented N-S (with a projected length
of $\sim$33,000 au). \citet{2016A&A...593A..49B} proposed that most of
the radio sources detected are part of the ionized jet and based on
their spectral indices, some of the emission of these knots is
non-thermal. The ionized jet is expanding at a velocity of $\sim 300$
km s$^{-1}$ and appears to be precessing as suggested by its S-shaped
morphology and the misalignment with respect to the larger scale CO
molecular outflow \citep{2016A&A...593A..49B}. Cores A and C are also
associated to UC/HC HII regions and are inside our \emph{SOMA} scale:
hence the higher flux densities measured at this scale. However, we
also measure a high flux density at our \emph{Intermediate} scale
compared with the \emph{Inner} scale.

\subsubsection{G45.47$+$00.05}

G45.47$+$00.05 appears as a resolved and slightly elongated radio
continuum source \citep[e.g.,][]{2009A&A...501..539U,
  2017ApJS..230...22T}, associated with at least one molecular outflow
roughly oriented with a N-S axis \citep{1996ApJ...462..339W}.
\citet*{1989ApJ...340..265W} observed this source at 6 cm with an
angular resolution of $\sim$0$\rlap.^{\prime \prime}$4 and cataloged
it as an irregular UC HII region, since their observations revealed a
NE-SW elongation as well as a slight NW-SE
elongation. \citet{1996ApJ...462..339W} studied the centimeter continuum
spectrum of G45.47$+$00.05 and suggested that it was consistent with a
partially ionized stellar wind. Additionally,
\citet{2017ApJS..230...22T} presented VLA 1.3 cm observations
estimating a flux density of $\sim$ 180 mJy with an angular resolution
of $\sim$1$^{\prime \prime}$ that also revealed an elongated morphology along
the NW-SE axis, suggesting an ionized jet nature. In this work we
present very high resolution ($\sim$0$\rlap.^{\prime \prime}$04) VLA
Q-band archival data where we can see for the first time (to the
knowledge of the authors) that G45.47$+$00.05 is fragmented into at
least two centimeter continuum sources $\sim$0$\rlap.^{\prime
  \prime}$4 apart (or $\sim$3400 au at the distance of
G45.47$+$00.05). The southern source, which we have named
G45.47$+$00.05S, is very bright at 43.9 GHz with a flux density of
$\sim$ 110 mJy, appears slightly elongated along the N-S axis and its
peak intensity position at that frequency is
RA(J2000)=$19^{h}14^{m}25{\rlap.}^{s}677$,
Dec(J2000)$=+$$11^{\circ}09^{\prime}25{\rlap.}^{\prime\prime}56$. The
northern source, which we have named G45.47$+$00.05N, is weaker at
43.9 GHz with a flux density of $\sim$ 20 mJy, has a jet-like
morphology elongated in the E-W axis and its peak intensity
position at that frequency is
RA(J2000)=$19^{h}14^{m}25{\rlap.}^{s}683$,
Dec(J2000)$=+$$11^{\circ}09^{\prime}25{\rlap.}^{\prime\prime}96 $. Our
\emph{Inner} scale is centered around G45.47$+$00.05S at Q-band where
the source is resolved.

\subsubsection{IRAS 20126$+$4104}

IRAS 20126$+$4104 has infrared emission that is slightly elongated
along the same NW-SE axis as the associated bipolar molecular outflow
\citep[e.g.,][]{1997A&A...325..725C}. \citet{2016ApJS..227...25R}
detected at least five centimeter continuum sources that are within our
SOMA aperture as seen in Figure \ref{fig:VLA_contours}.
\citet{2007A&A...465..197H} suggested that the central, slightly
elongated sources seen in Figure \ref{fig:VLA_contours} correspond to
two collimated ionized jets, one of them composed of radio sources N1
and N2 and the second ionized jet known as source S. It has been
suggested that source N1 sits at the origin of the larger scale
molecular outflow and also appears to be surrounded by a nearly
Keplerian and stable accretion disk
\citep[e.g.,][]{2014A&A...566A..73C,2016ApJ...823..125C}. Additionally,
\citet{2007A&A...465..197H} reported that the southern radio source
towards IRAS 20126$+$4104 (labeled I20var) is a variable radio source,
consistent with gyrosynchrotron emission from a T-Tauri star. I20var is
located  $\sim$5$\rlap.^{\prime \prime}$0
to the SE of source N1. The other two sources found by \citet{2016ApJS..227...25R} were
labeled source C (located $\sim$4$\rlap.^{\prime \prime}$0 NW apart
from N1) and G78.123$+$3.629 (located $\sim$11$\rlap.^{\prime
  \prime}$5 SE apart from N1, very near to the edge of the millimeter
dust core) and are new detections with reported spectral indices that
are consistent with non-thermal emission. It is possible that at the
distance of IRAS 20126$+$4104 the VLA could still detect low-mass
stars (Rosero et al. in prep), thus these non-thermal highly variable
radio sources could be flaring T-Tauri stars. In this scenario, IRAS
20126$+$4104 would be a small multiple system of two high-mass
protostars surrounded by at least three pre-main sequence low-mass
stars. The four central sources are within a projected radius of
$\sim$4$\rlap.^{\prime \prime}$5 (7400 au at the distance of IRAS
20126$+$4104).

\subsubsection{Cepheus A}

Cepheus A is a well known star-forming region and it is seen in the
infrared as an elongated source oriented NE-SW. This region is the
host of a complex multipolar CO molecular outflow that is likely
powered by a small cluster of deeply embedded protostars
\citep[e.g.,][]{2006ApJ...638..878C,2013arXiv1305.4084Z}. Our SOMA
aperture encloses the whole Cepheus A East region that is composed of
several centimeter continuum sources as seen in
\citet{2013arXiv1305.4084Z}. The brighter centimeter continuum source
found towards this region, known as HW2, is an ionized thermal jet
oriented NE-SW and it is thought to be the main driver of the complex
outflow activity in Cepheus A and to contribute at least half of the
total luminosity of the region
\citep[e.g.,][]{2013arXiv1305.4084Z}. Also, the ionized jet in HW2 is
associated and aligned with a bipolar HCO$^{+}$ molecular outflow
\citep{1999ApJ...514..287G}. \citet{2006ApJ...638..878C} resolved the
Cepheus A HW2 ionized jet at 3.6 cm in three radio sources, where HW2
is the central powering source and the NE and SW sources are knots
along the jet that are moving away from HW2 at a velocity of $\sim$500
km s$^{-1}$. Based on maser and millimeter observations some authors
have suggested that HW2 has an associated circumstellar disk
\citep[e.g.,][]{2005Natur.437..109P, 2014A&A...562A..82S,
  2017A&A...603A..94S}. We used the reported values from
\citet{2006ApJ...638..878C} with angular resolutions of
$\sim$0$\rlap.^{\prime \prime}$27 and $\sim$0$\rlap.^{\prime
  \prime}$05 for the central source for our study of the \emph{Inner}
scale.  We used the reported values from \citet{1994ApJ...430L..65R}
of the radio jet for our study of the \emph{Intermediate} scale. We
are not analyzing the SOMA aperture towards this region due to a lack
of a complete suit of observations that enclosed all the radio sources
at this scale.

\subsubsection{NGC 7538 IRS9}

NGC 7538 IRS9 is part of a cluster of infrared regions that host
several high-mass star forming cores.
\citet{2005ApJ...621..839S} detected a
faint and marginally resolved radio source (known as IRS 9) in this
region using VLA data at angular resolution of $\sim$1$^{\prime
  \prime}$ at 3.6 cm and 6 cm and with a rms noise of 60 $\mu$Jy
beam$^{-1}$ and 100 $\mu$Jy beam$^{-1}$, respectively. Based on their
measured spectral index they have suggested that IRS 9 is free-free
emission from an ionized jet, a scenario that is consistent with the
observed collimated, compact and bipolar HCO$^{+}$ molecular outflow
that is centered on IRS 9 and oriented
E-W. \citet{2005A&A...437..947V} detected IRS 9 using VLA data at 43.3
GHz at two angular resolutions of $\sim$0$\rlap.^{\prime \prime}$5 and
$\sim$0$\rlap.^{\prime \prime}$05. Our results show further evidence
that the radio emission from IRS 9 is dominated by free-free
emission. Additionally, from these archival observations we have
detected a new weak and unresolved radio source at 5.8 GHz, located
$\sim$1$\rlap.^{\prime \prime}$8 NE from IRS9 (or $\sim$4800 au). This
new radio detection has a flux density at 5.8 GHz of $\sim$ 0.2 mJy,
and its peak intensity position is
RA(J2000)=$23^{h}14^{m}01{\rlap.}^{s}844$,
Dec(J2000)$=+$$61^{\circ}27^{\prime}21{\rlap.}^{\prime\prime}46$. This
source appears to be undetected by the observations of
\citet{2005ApJ...621..839S}, which suggests that this newly detected
radio source may be a radio variable source or that their observations lacked the angular resolution to distinguish these two sources. We estimate an upper
limit on the radio spectrum for this new component which is consistent
with a flat or even a negative spectral index.

\newpage
\subsection{Radio SEDs}\label{spectral_indices}

In Figure \ref{fig:radio_SEDs} we present the centimeter SEDs for our
regions. The dashed lines are the best fit to the data of a power law
of the form $S_{\nu} \propto \nu^{\alpha}$, where $\alpha$ is the
spectral index, derived at the different scales: \emph{``SOMA''},
\emph{``Intermediate''} and \emph{``Inner''}, as described above. The
spectral index was calculated using the flux density at the central
frequencies from the images (and/or the available data in the
literature), thus $\alpha$ is calculated over a wide frequency range
($>$20 GHz), except for IRAS 07299$-$1651 where we only have data at  6
cm.  The presented archival data were observed at different angular
resolutions at different frequencies, with the lower frequency observed
(C-band: 6 cm) at a resolution around 10 times lower than the higher
frequency (Q-band: 0.7 cm). This is not a major concern for sources measured
to be very compact or unresolved at C-band because they will then be smaller than the largest angular scale for which the Q-band observations are sensitive. For the sources in this study (typically at about 2 kpc) this corresponds to a linear size of $\sim 2400$ au.

However, more extended sources may suffer from resolution bias 
and/or lack of short spacing data (resolved out), affecting our ability to recover a source's entire flux. 
Additionally, at the higher frequencies (K-band and
Q-band in our study) the fluxes are most likely measuring the
combination of dust and free-free emission (see
\citealt{2016ApJ...832..187B}). We assume that the fluxes at Q-band
are an upper limit on the free-free emission contribution.  The
uncertainty in the spectral index was calculated with a Monte Carlo
simulation that bootstraped the flux density uncertainties. We
estimated an upper limit in the spectral index for non-detections at
higher frequencies using a value of $S_{\nu}$ of 3$\sigma$.

\begin{figure}[htbp]
\centering
\begin{tabular}{cc}
\hspace*{\fill}%
\includegraphics[width=0.48\textwidth,  trim = 20 20 20 15, clip, angle = 0]{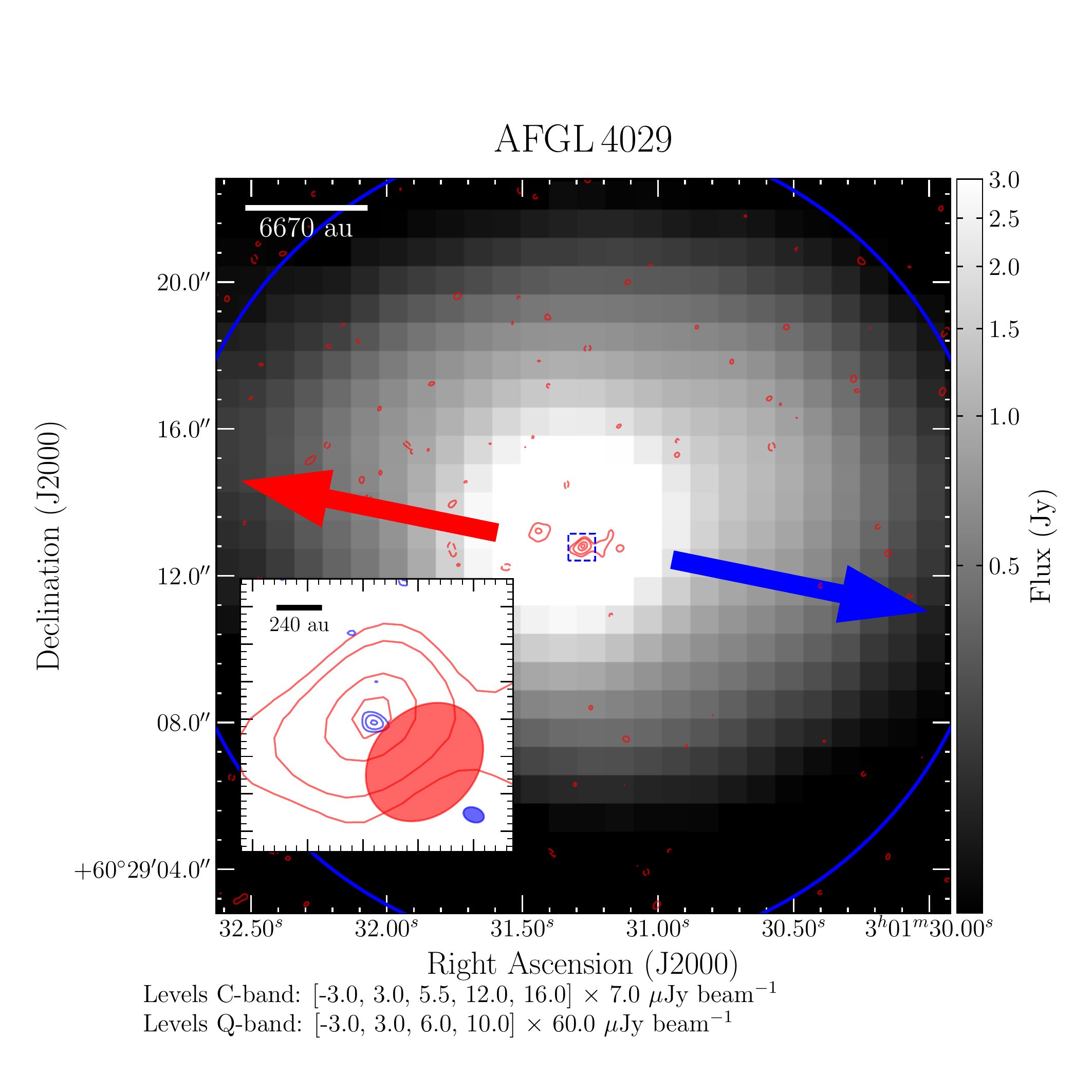}\hspace{-0.1cm} &
\includegraphics[width=0.48\textwidth,  trim = 20 20 20 15, clip, angle = 0]{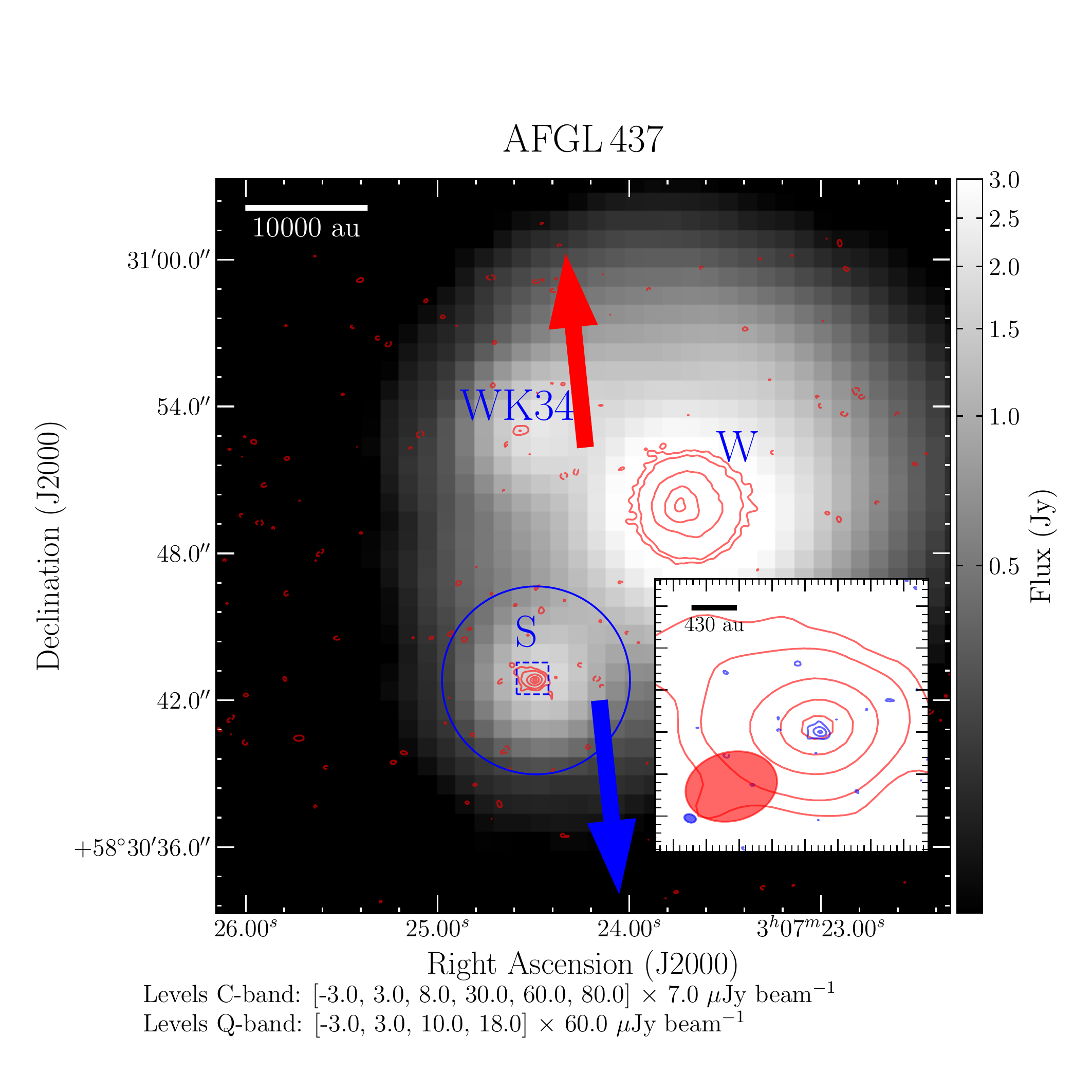}\\
\includegraphics[width=0.48\textwidth,  trim = 20 10 20 15, clip, angle = 0]{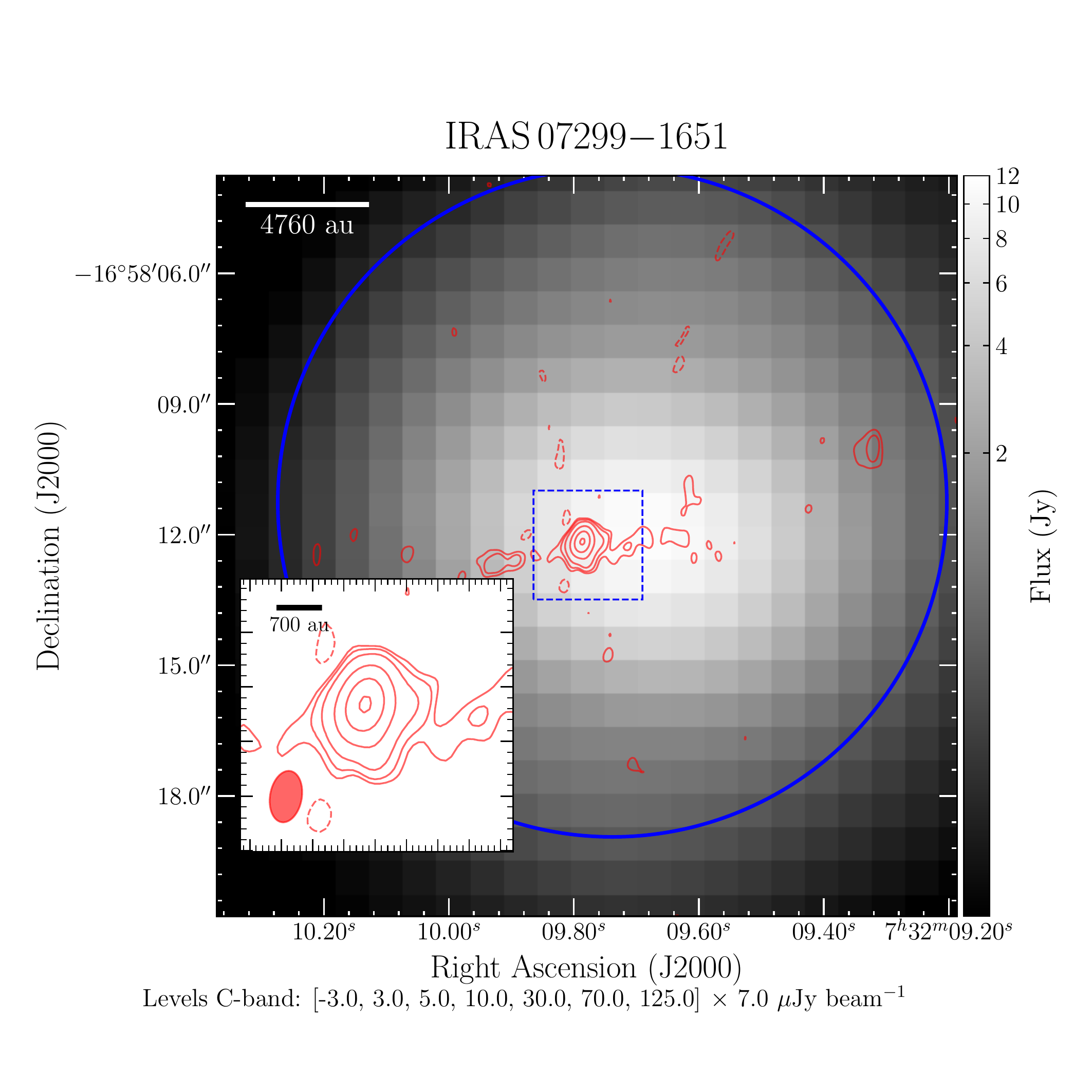}\hspace{-0.1cm} &
\includegraphics[width=0.48\textwidth,  trim = 20 10 20 50, clip, angle = 0]{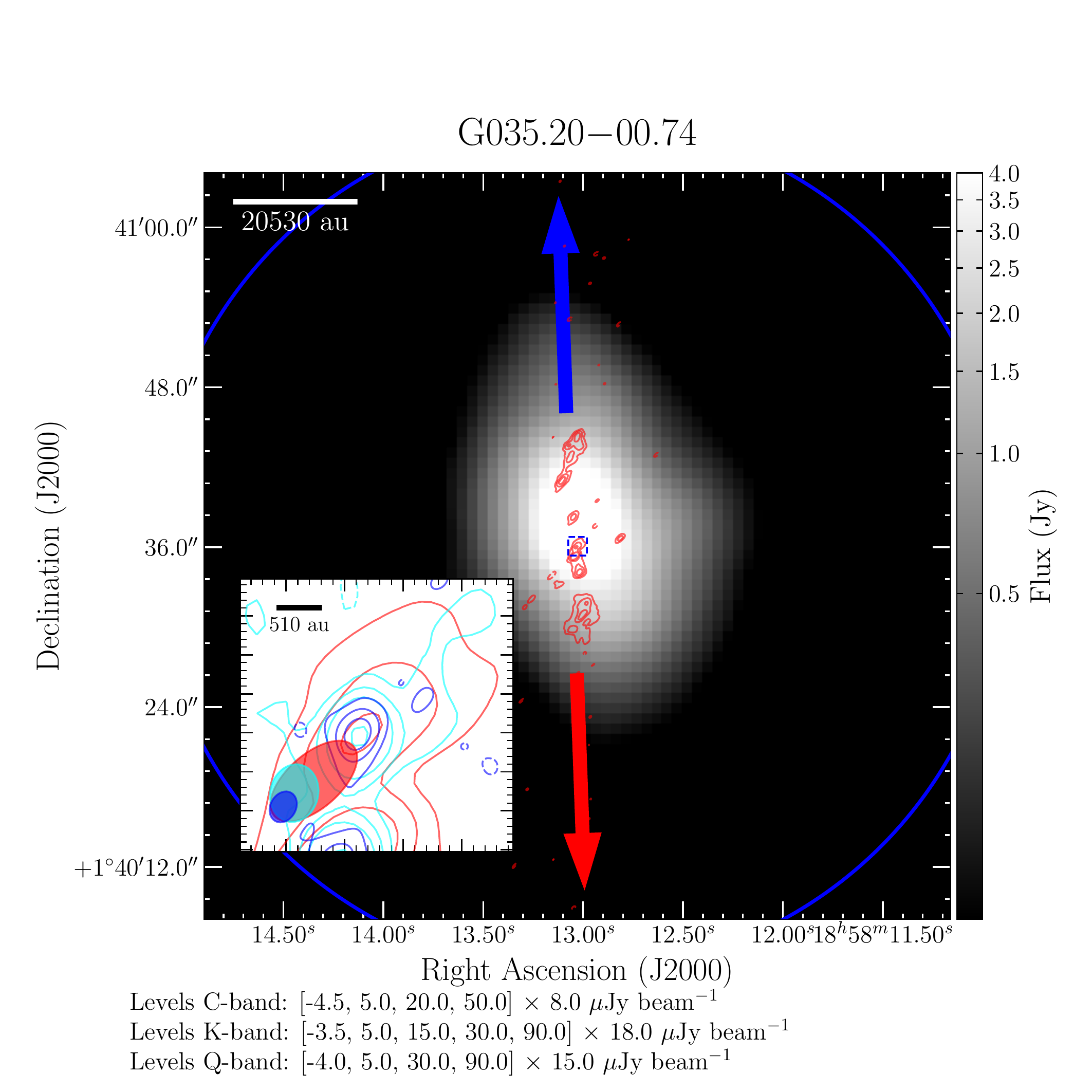}\\

\end{tabular}

\caption{\small{
Images are SOFIA-FORCAST 37$\mu$m with VLA contours --red: C-band (6 cm);
magenta: K-band (1.3 cm); blue: Q-band (0.7 cm)--
of the combined radio maps overlaid. Contours in Cepheus A are 8.3 GHz
and 42.9 GHz from \citet[][ see their Figure 2 for a higher resolution image of these contours]{2006ApJ...638..878C}  and the location and size of this inset are represented by the small blue box shown towards the region. The dashed squares
correspond to the area of the inset image showing a zoom in of the
central region and the synthesized beams are shown in the lower
corners of these insets. The blue circles are the SOMA apertures used
by \citetalias{2017ApJ...843...33D} and reported in their Table~2
(aperture radius defined from 70 $\mu$m emission except for
IRAS$\,$07299$-$1651 where it is set at 37.1 $\mu$m).  The blue and red arrows represent the direction of a molecular outflow detected towards the region.} A scale bar in
units of au is shown in the upper left of the figures.  }
 \label{fig:VLA_contours}
\end{figure}

\begin{figure}[htbp]
\centering
\begin{tabular}{cc}
  \ContinuedFloat
\hspace*{\fill}%
\includegraphics[width=0.48\textwidth,  trim = 20 20 20 15, clip, angle = 0]{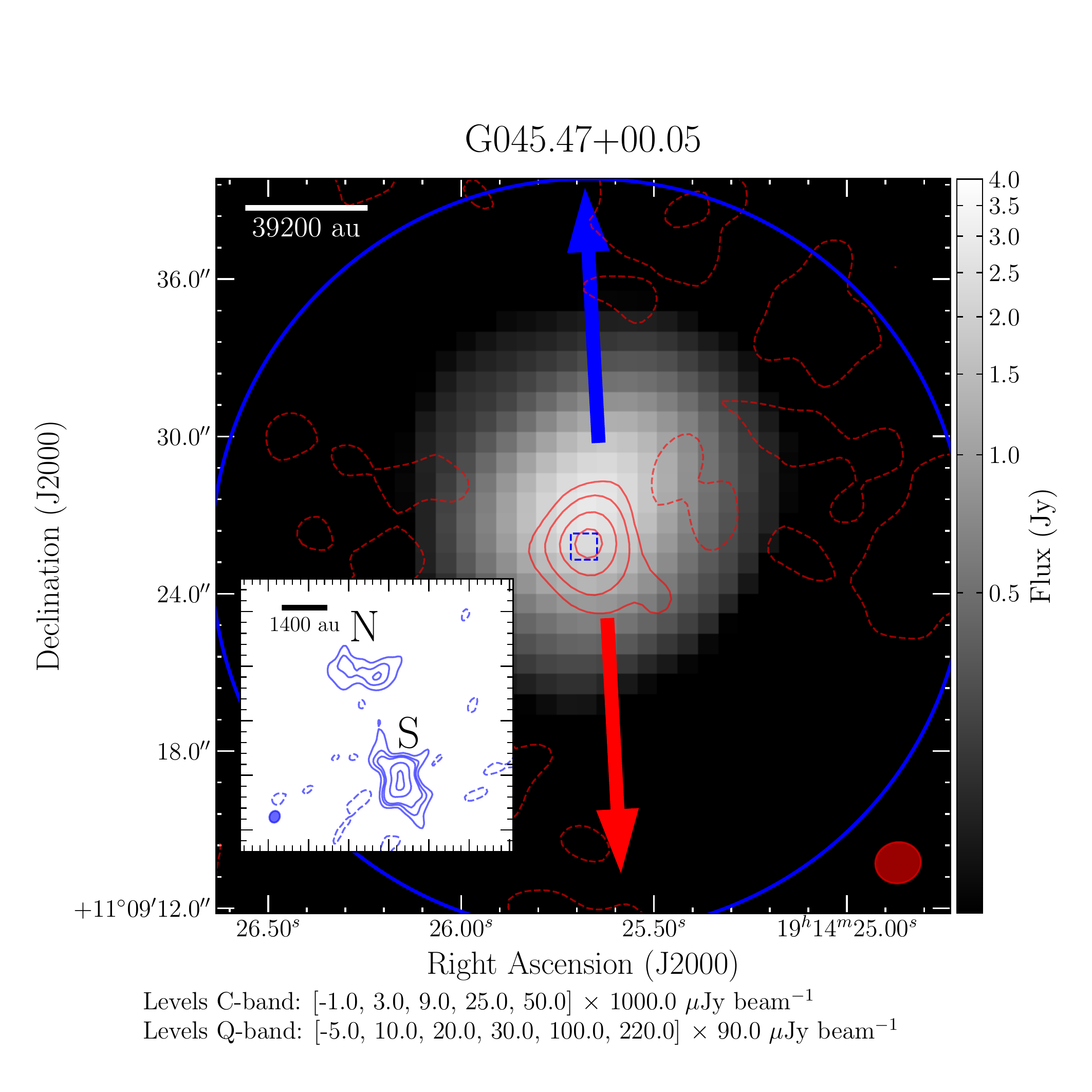}\hspace{-0.1cm} &
\includegraphics[width=0.48\textwidth,  trim = 20 20 20 15, clip, angle = 0]{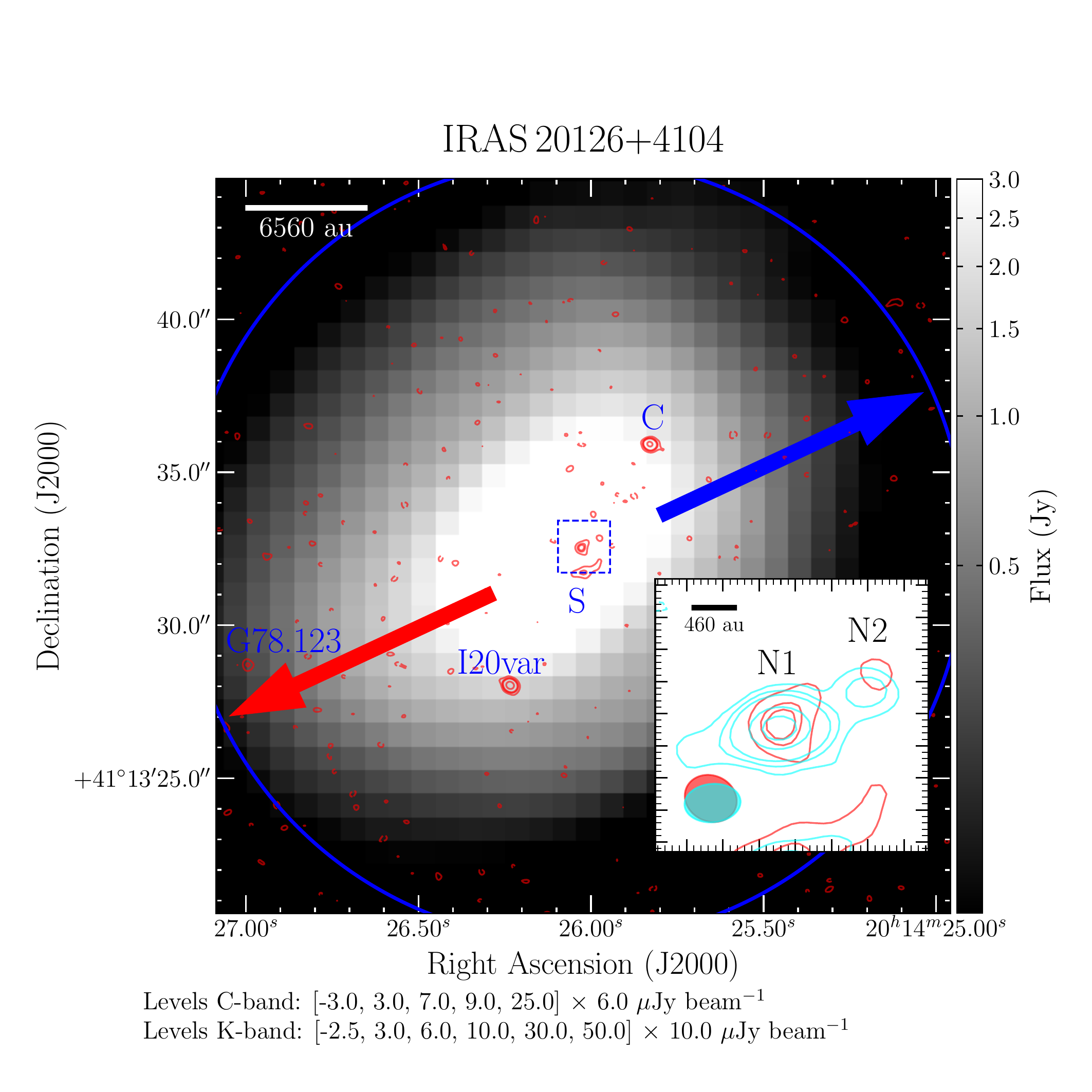}\\
\raisebox{0.07\height}{\includegraphics[width=0.55\textwidth, trim = 30 20 20 15, clip, angle = 0]{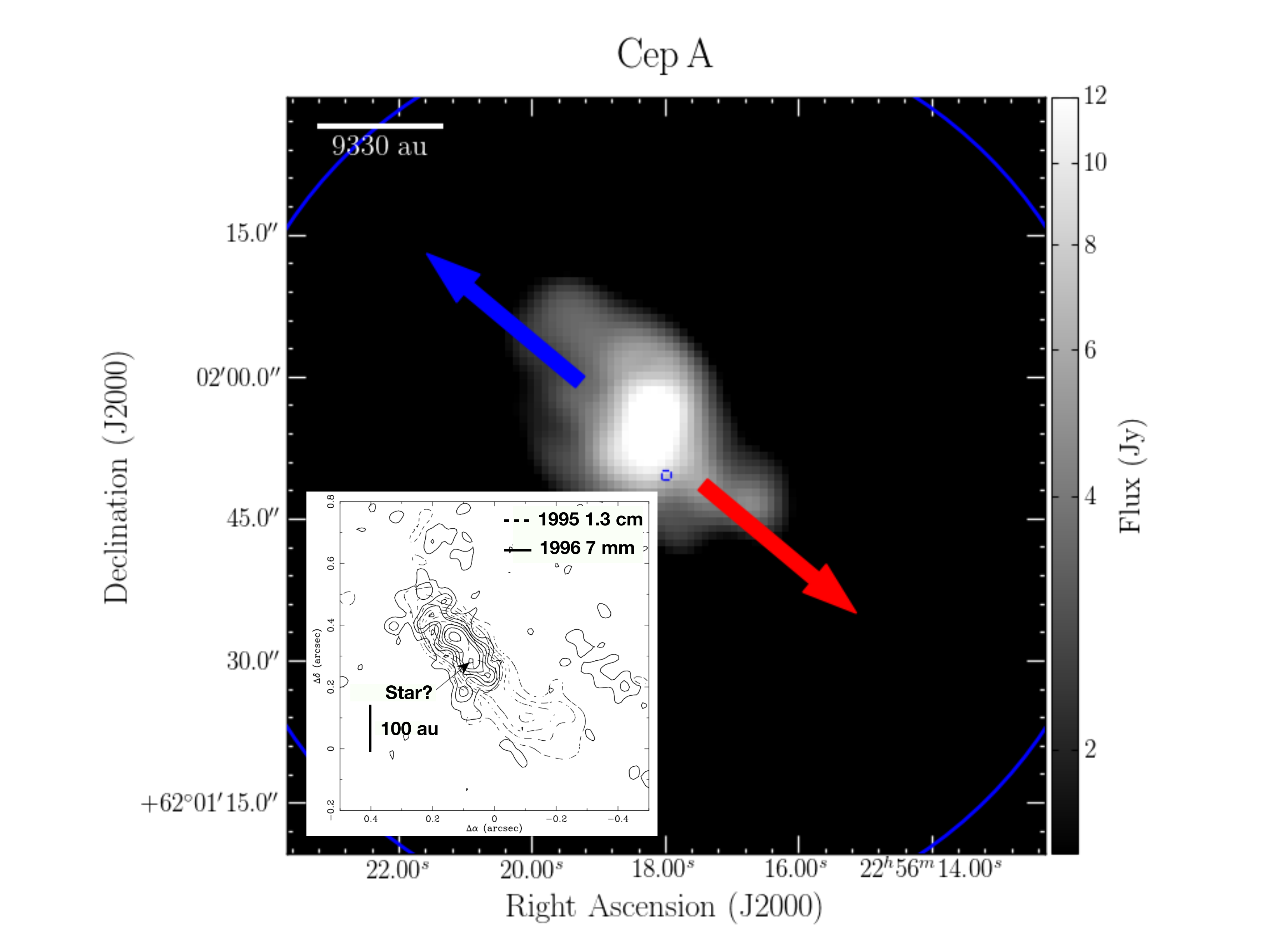}}\hspace{-0.6cm} &
\includegraphics[width=0.48\textwidth,  trim = 20 20 20 15, clip, angle = 0]{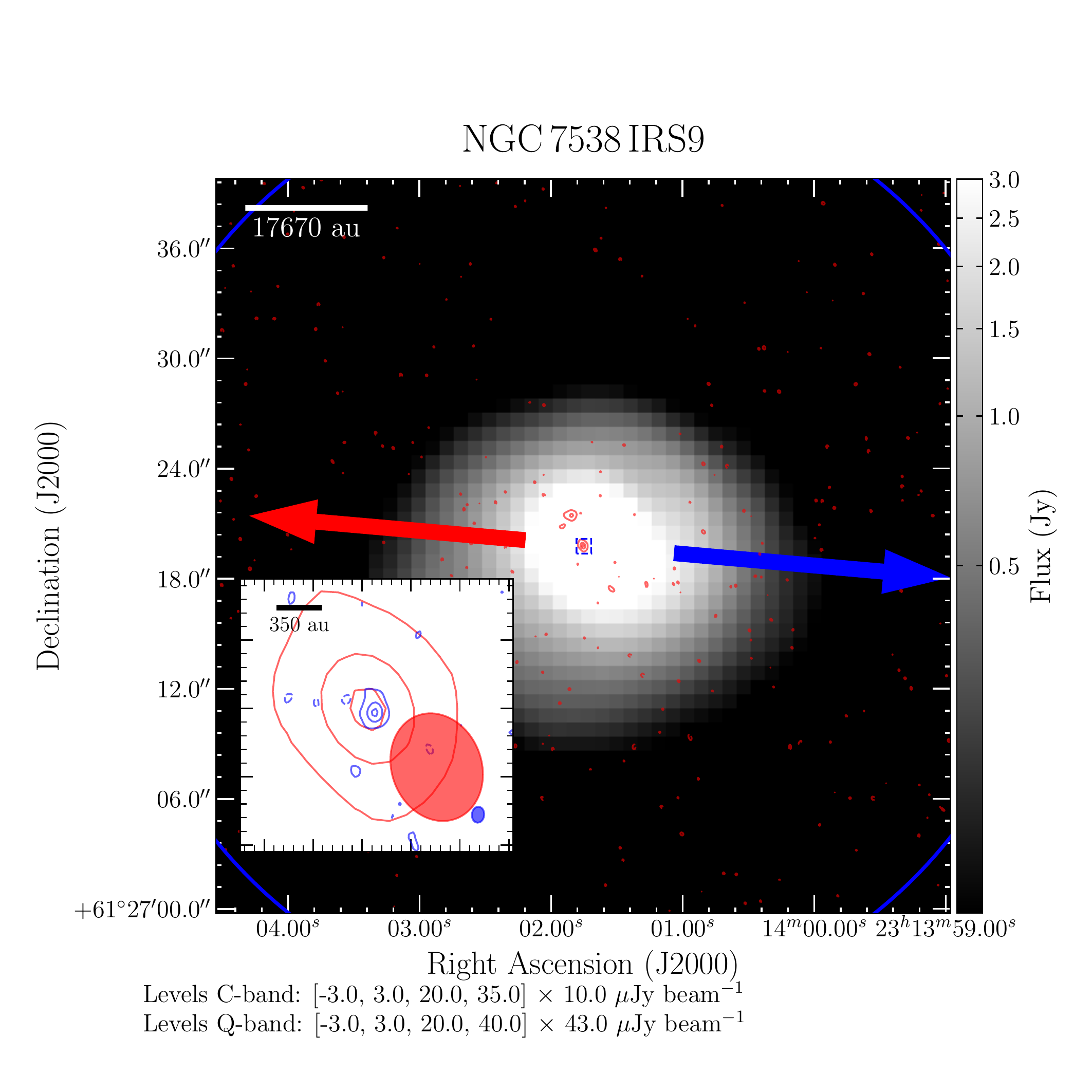}\\

\end{tabular}

 \hspace*{\fill}%
\caption{\small{Continued.}}
 \label{fig:VLA_contours}
\end{figure}

\begin{figure}[htbp]
\centering
\begin{tabular}{cc}
\hspace*{\fill}%
\includegraphics[width=0.48\textwidth,  trim = 20 20 20 15, clip, angle = 0]{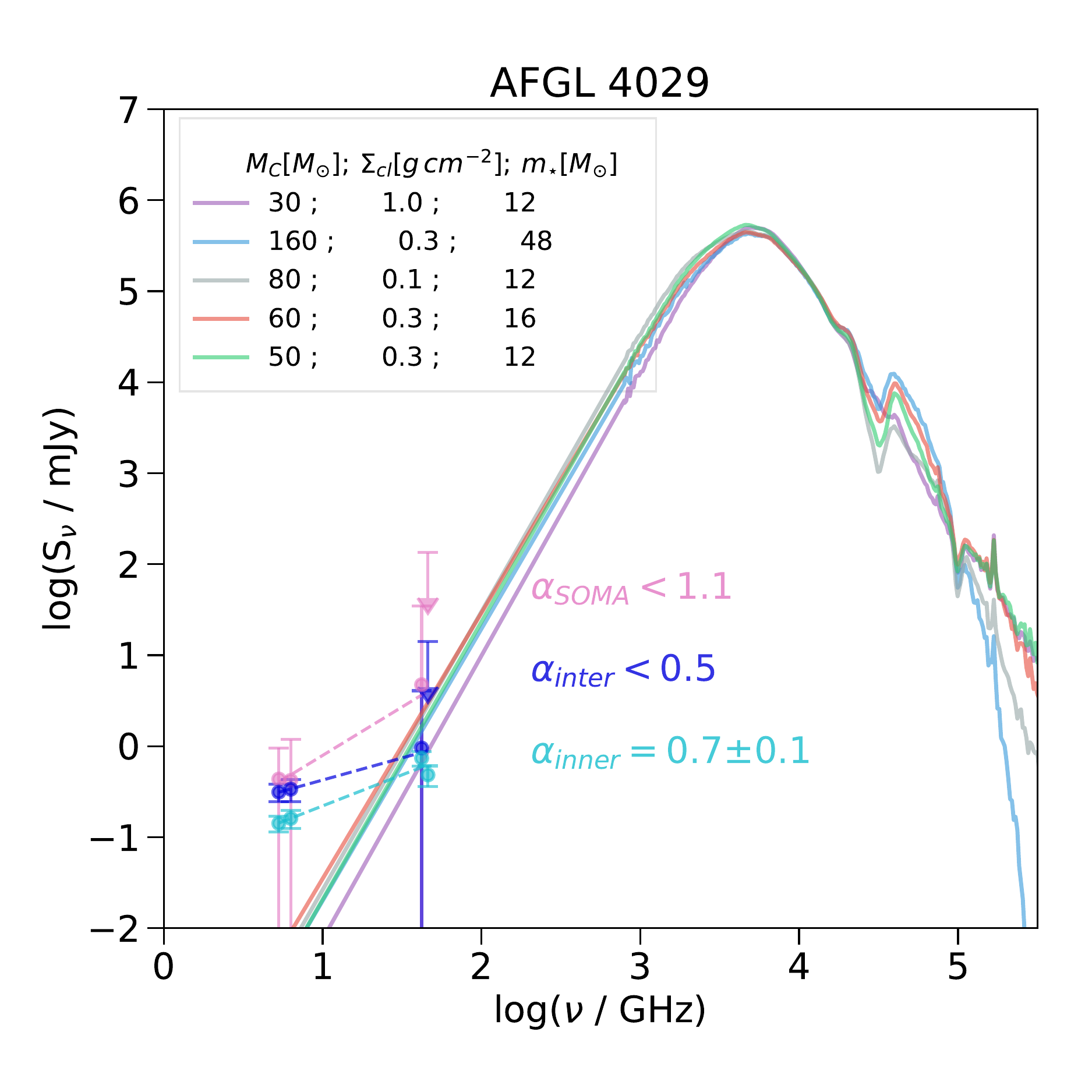}\hspace{-0.1cm} &
\includegraphics[width=0.48\textwidth,  trim = 20 20 20 15, clip, angle = 0]{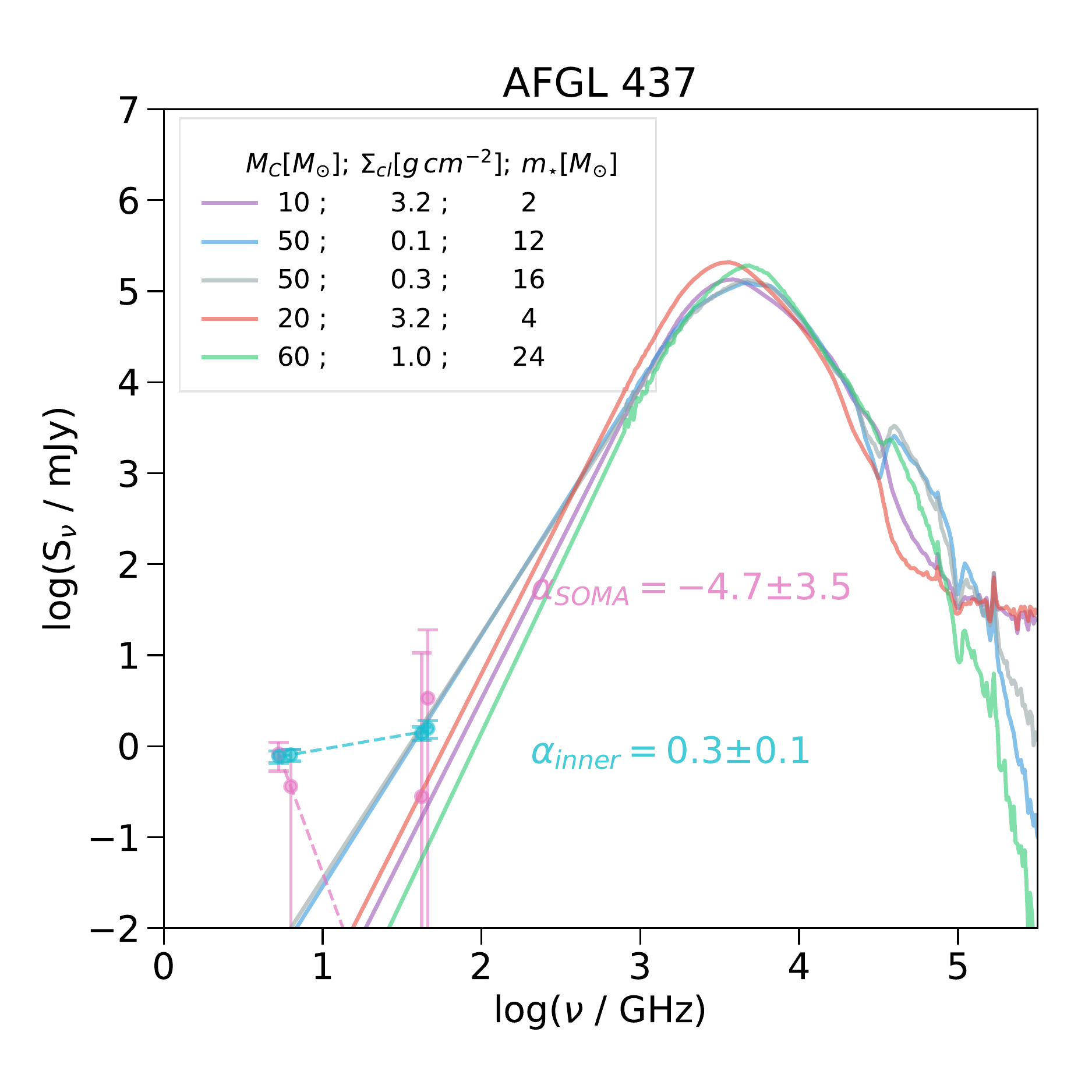}\\
\includegraphics[width=0.48\textwidth,  trim = 20 20 20 15, clip, angle = 0]{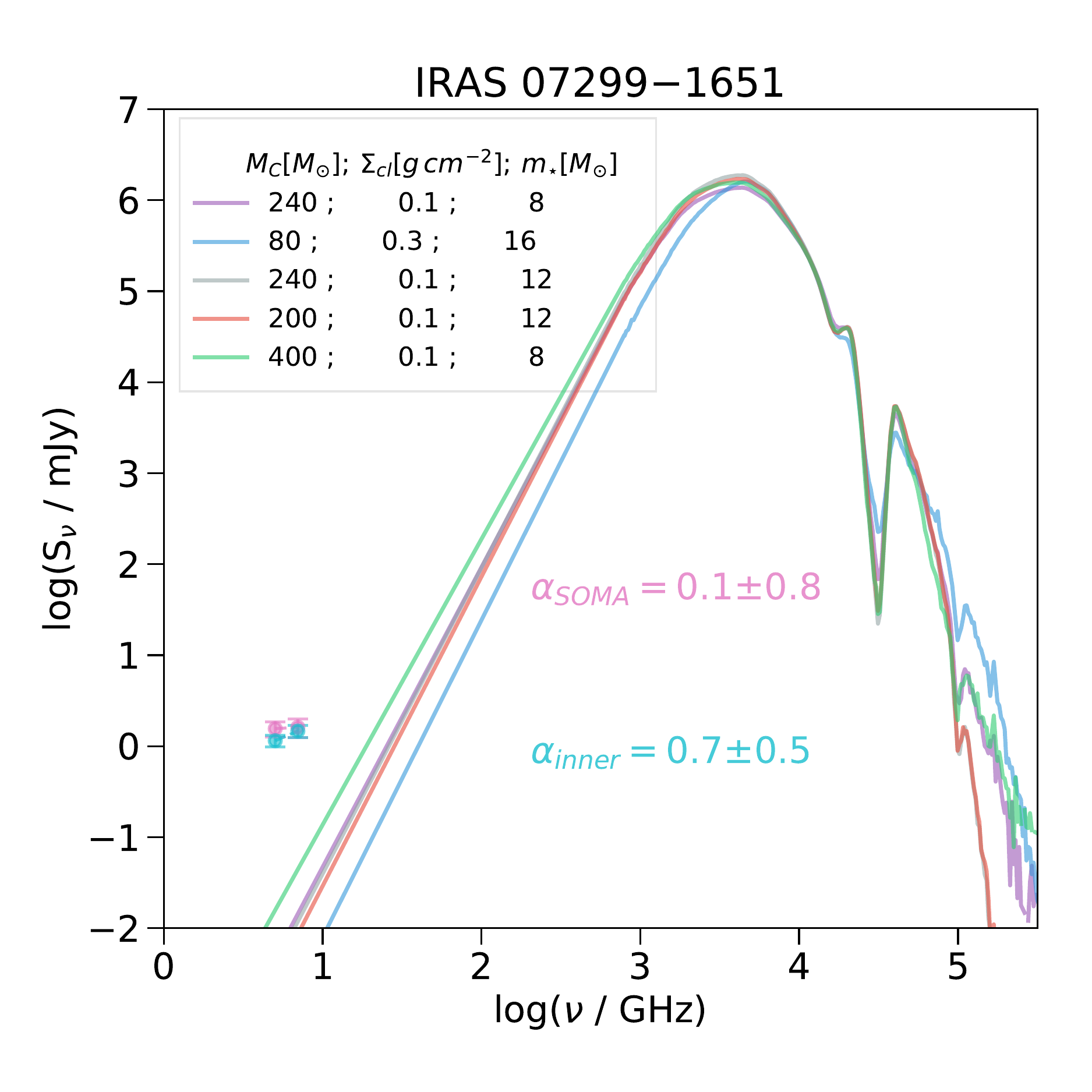}\hspace{-0.1cm} &
\includegraphics[width=0.48\textwidth,  trim = 20 20 20 15, clip, angle = 0]{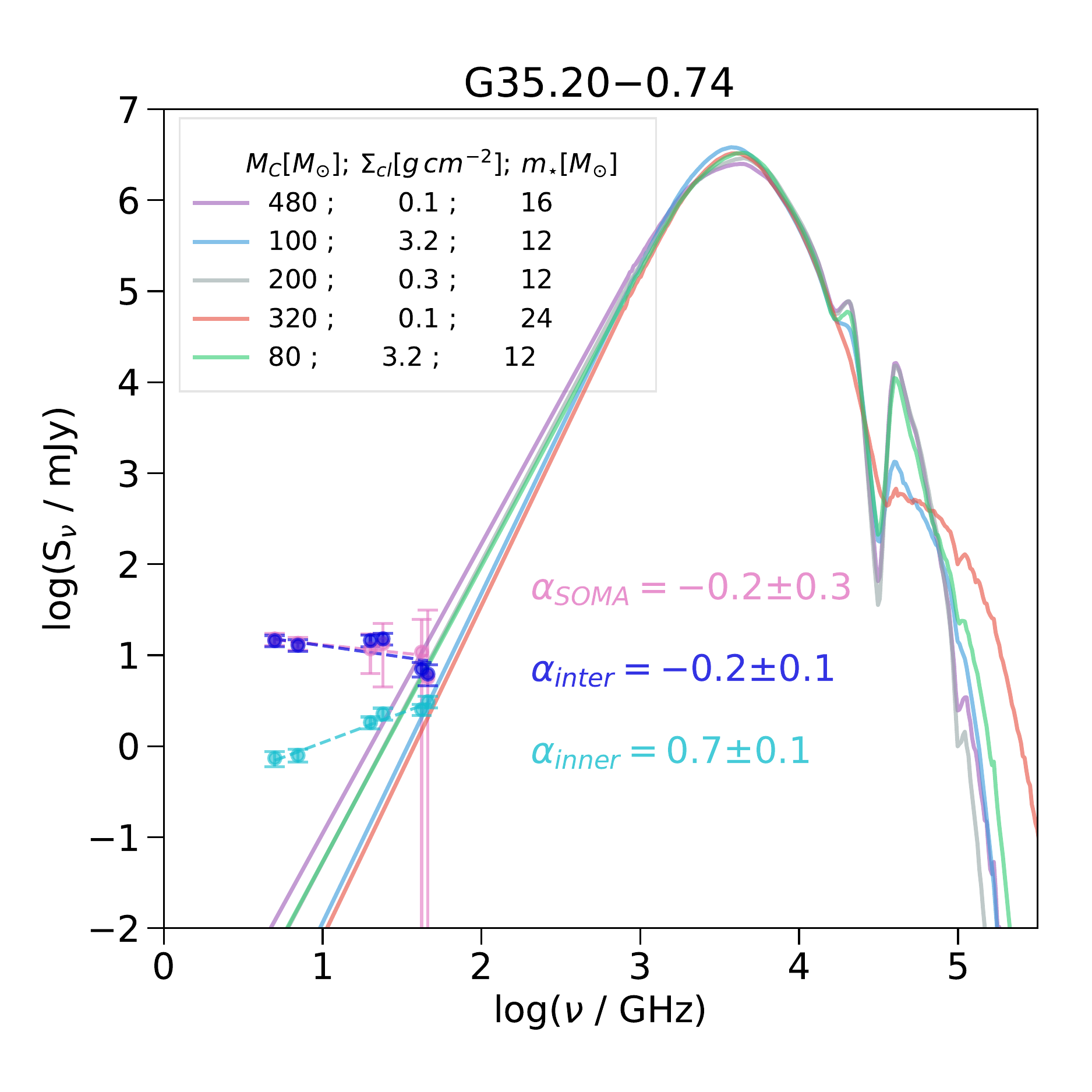}\\

\end{tabular}
\caption{\small{
Observed radio spectral energy distributions of SOMA protostars. The
circles correspond to the flux density as a function of frequency for
each scale (magenta: SOMA; blue: Intermediate; cyan: Inner). Error
bars are explained in \S\ref{spectral_indices}. The dashed lines
are the best fit to the data from a power law of the form $S_{\nu}
\propto \nu^{\alpha}$. The solid lines show the best five IR SED
ZT models as fit by \citetalias{2017ApJ...843...33D} (see legend).}}
\label{fig:radio_SEDs}
\end{figure}

\begin{figure}[htbp]
\centering
\begin{tabular}{cc}
  \ContinuedFloat
\hspace*{\fill}%
\includegraphics[width=0.48\textwidth,  trim = 20 20 20 15, clip, angle = 0]{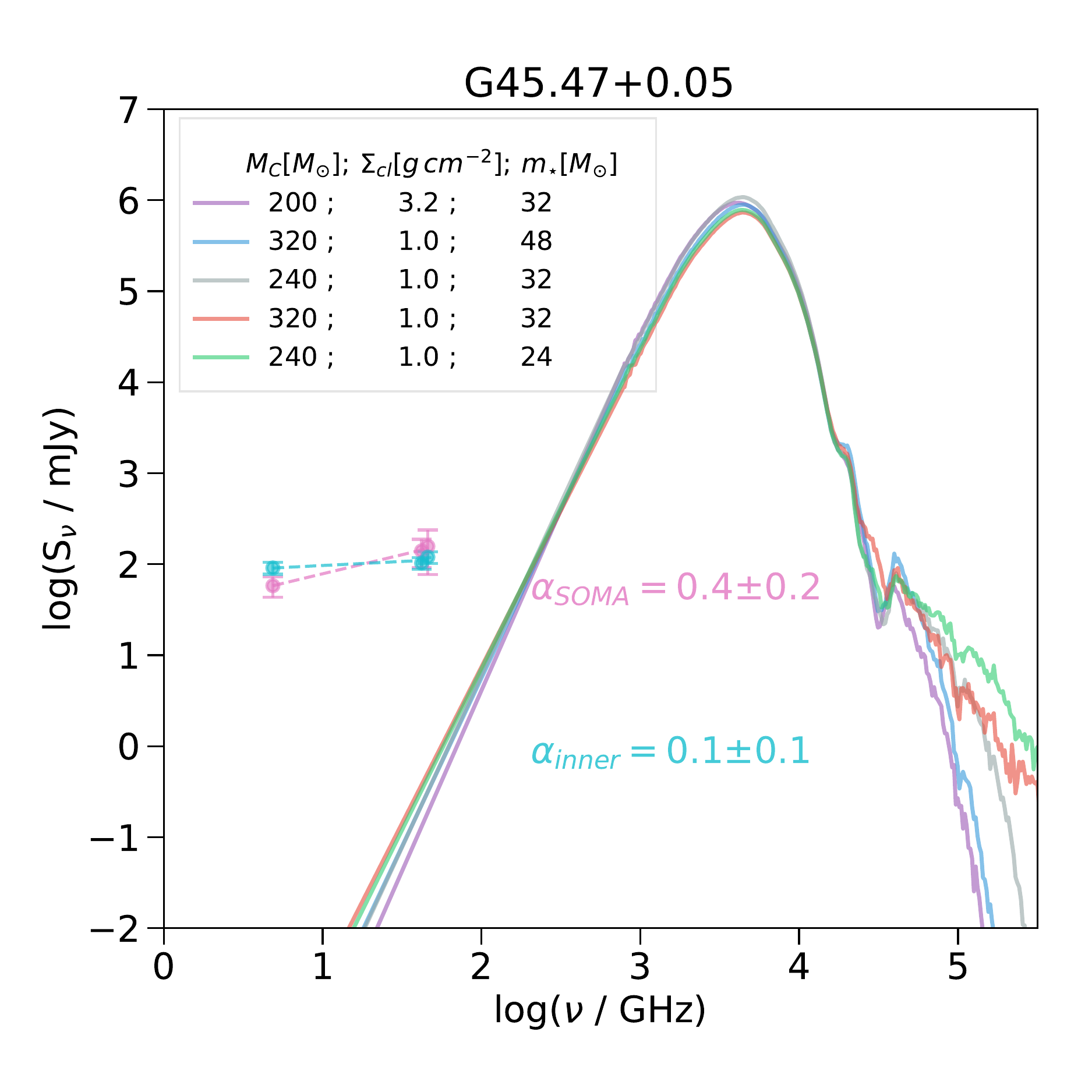}\hspace{-0.1cm} &
\includegraphics[width=0.48\textwidth,  trim = 20 20 20 15, clip, angle = 0]{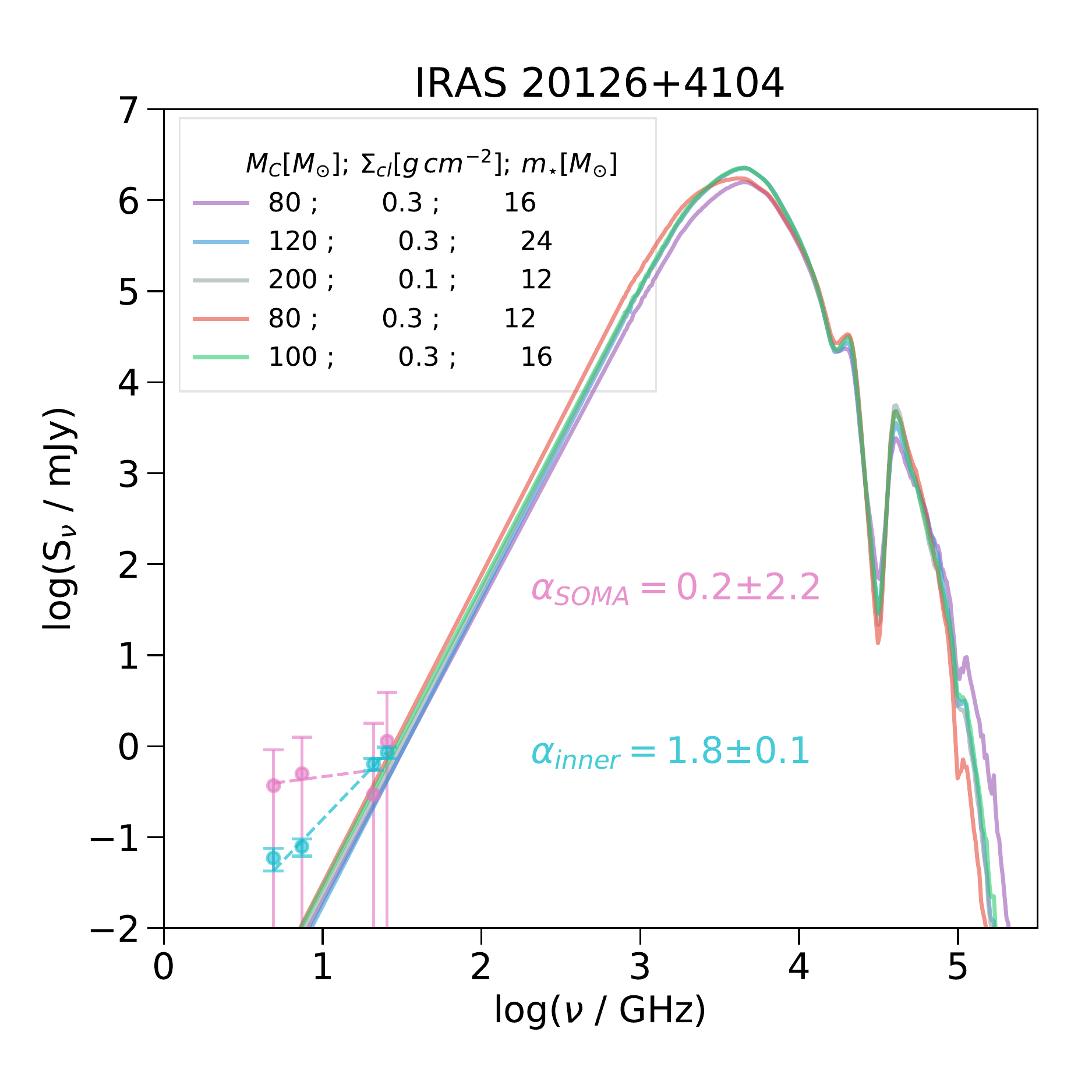}\\
\includegraphics[width=0.48\textwidth,  trim = 20 20 20 15,  clip, angle = 0]{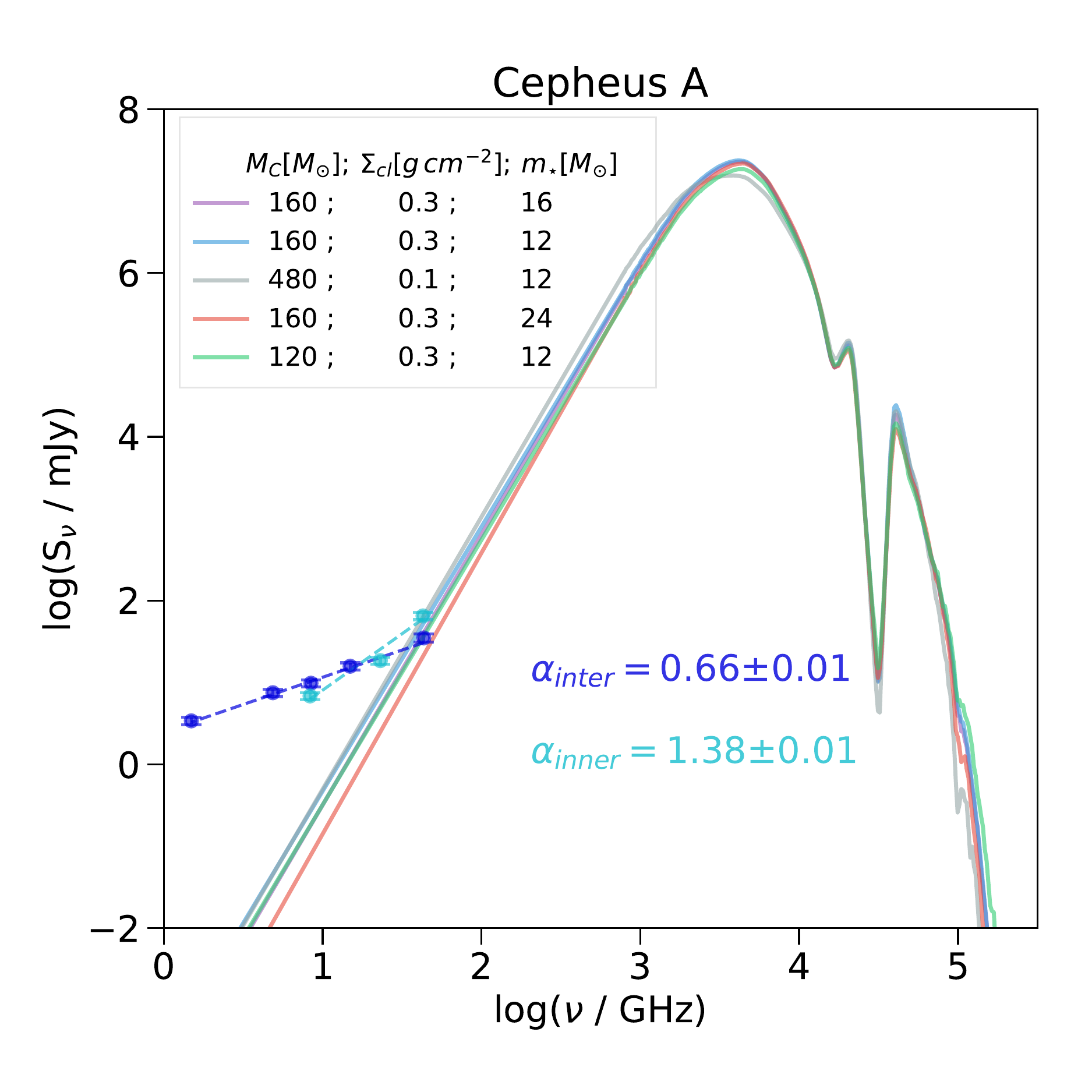}\hspace{-0.1cm} &
\includegraphics[width=0.48\textwidth,  trim = 20 20 20 15, clip, angle = 0]{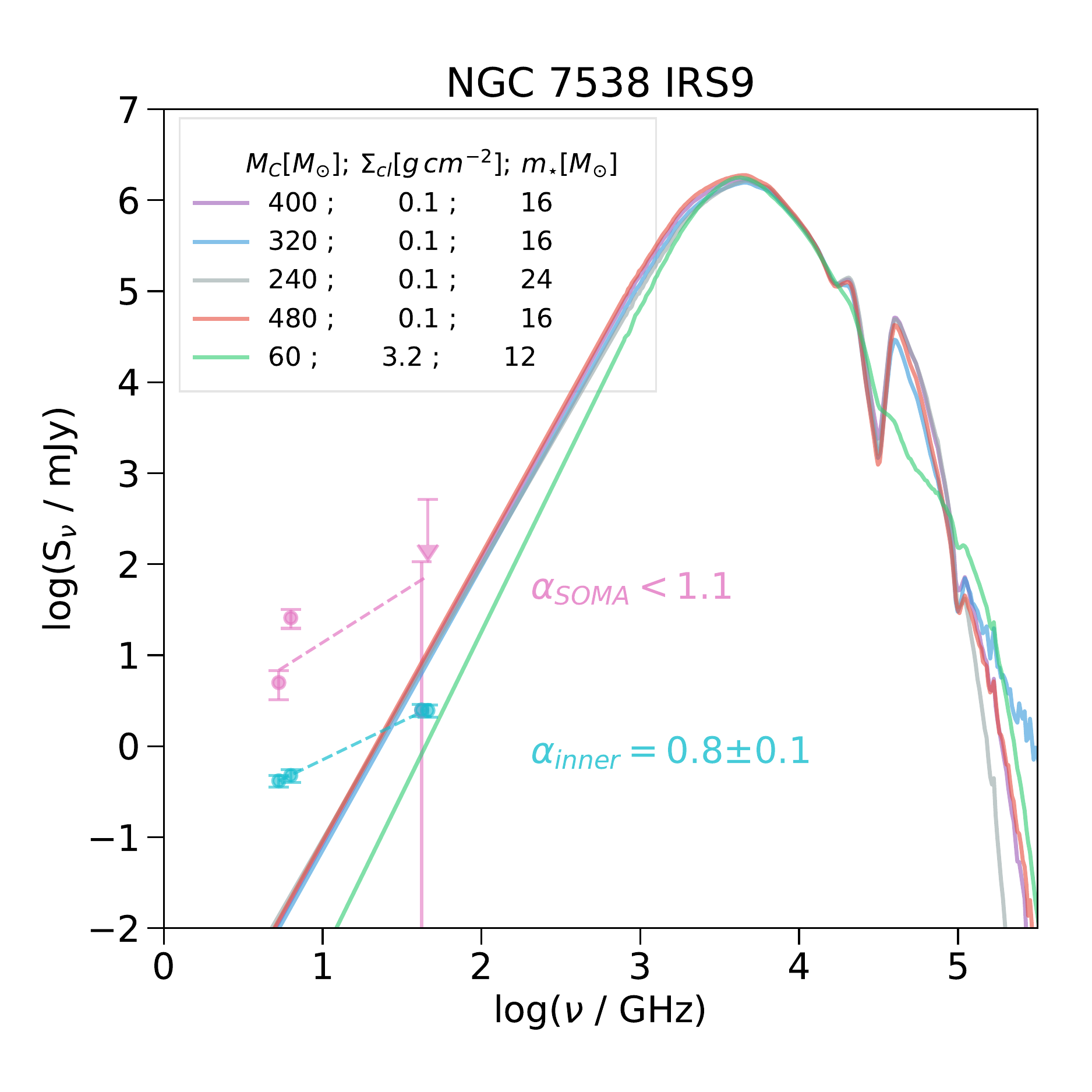}\\
\end{tabular}

 \hspace*{\fill}%
\caption{\small{Continued.}}
 \label{fig:radio_SEDs}
\end{figure}

\section{Analysis} \label{sec:models}

The centimeter continuum emission associated with the regions located
at distances $>$1 kpc have low radio fluxes ($<$3 mJy) in our
\emph{Inner} scale, except for G45.47$+$0.05. The analyzed data do not
allow us to make a systematic study of the nature of this detected
emission, but we favor the speculation from previous studies that some
of them are part of a protostellar jet/wind, at least based on
their morphology, as in the case of AFGL 4029, G35.20$-$0.74, IRAS
20126$+$4104. Cepheus A is the only region in this study located at a
distance $<$1 kpc and it is one of the best examples of a collimated
ionized jet from a high-mass protostar in the literature
\citep{1994ApJ...430L..65R}.

Figure \ref{fig:SEDs} shows the extended spectral energy
distributions, i.e., including radio fluxes as well as the infrared
fluxes from \citetalias{2017ApJ...843...33D}, for our eight
sources. The solid colored lines (except yellow ones) correspond to
the best five models obtained from the ZT protostellar radiative
transfer models. As explained in \citetalias{2017ApJ...843...33D}, the
data at $\lesssim$8 $\mu$m are considered to be upper limits since PAH
emission and transiently heated small grain emission are not well
treated in the ZT models.

In the cases of AFGL 4029 and IRAS 20126$+$4104, all of the best five
models have a higher predicted radio flux than what is observed. In
these cases, we analytically select alternative models within the top
20 models from the IR fitting and that are expected to have lower
radio fluxes. In general, the free-free flux quickly increases when
the photoionized region breaks out the inner disk wind. Therefore, we
selected alternative models that are expected to be just before the
break-out phase, which is typically at the protostellar mass of
$m_*\simeq 10 \: (\dot{m_*}/10^{-4}
\:M_\odot{\rm\:yr}^{-1})^{0.28}\:M_\odot$
(\citetalias{2016ApJ...818...52T};
\citealt{2017ApJ...849..133T}). Ultimately, as in the ZT model for IR
and sub-mm wavelengths, we aim to prepare a full suite of model grids
for centimeter wavelengths, but since this is computationally
expensive we defer it to a future paper.

The parameters of the best-fit ZT models are listed in Table
\ref{model_parameters} but now ordered from best to worst as measured
by the reduced $\chi^{2}$ from the inner scale
($\chi^{2}_{all\_inner}$). This is because the ZT and
\citetalias{2016ApJ...818...52T} models are developed assuming a
single protostar within a core and with a focus on the inner
regions. The reduced $\chi^{2}$ is estimated using equation 4 from
\citet{2018ApJ...853...18Z}. The radio data occurring within the same band (i.e., with very similar frequencies) were averaged together to give more equal weight over the SED.

\subsection{Comparing models with data}

We now describe the results of the \citetalias{2016ApJ...818...52T}
models for each of the eight regions using the best five models from
the ZT grid as examples, except for AFGL 4029 and IRAS 20126$+$4104
where we had to identified a best overall model from the ZT results
that is beyond the best five (but within the best 20 results). We
center our attention on the results from the \emph{Inner} scale since
the \citetalias{2016ApJ...818...52T} model assumes that a prestellar
core collapses to form a single high-mass star and it is mostly
focussed on the inner regions.

\emph{AFGL 4029:}\, The best two fit models using MIR to FIR data
alone to build the SED have $\chi^{2}_{IR} =$ 1.07 and 1.16 and have
parameters of protostellar mass $m_{\star} =$ 12 and 48 $M_{\sun}$,
core mass $M_{c} =$ 30 and 160 $M_{\sun}$ and $\Sigma_{\rm cl} =$ 0.3 and
1 g\,cm$^{-2}$ clumps, respectively. Including the centimeter continuum
fluxes to better cover the SED we can see in Figure \ref{fig:SEDs}
that the best five models seem to be overestimating the expected
free-free emission by $\sim$1 dex. Therefore, after exploring other
resulting models from the \citetalias{2017ApJ...843...33D} study, we
found a matching model with $\chi^{2}_{IR}$ = 1.74 and
$\chi^{2}_{all\_inner} = $11.07, which yields protostellar parameters
of $m_{\star} = 8\,M_{\sun}$, a relatively low mass of the core ($M_{c} =\,80 M_{\sun}$) and
$\Sigma_{cl}=$ 0.1 g\,cm$^{2}$,
accreting at $5\times 10^{-5} M_{\sun}\,yr^{-1}$ and a bolometric
luminosity of $\sim 10^{4} L_{\sun}$.  

\emph{AFGL 437S:}\, For source AFGL 437S we lack reliable FIR
measurements at wavelengths $>$40 $\mu$m (see above) and we have only
three effective data points (plus the 3--8 $\mu$m data treated as
upper limits), thus the results from the ZT models in this case are
not that well constrained. In this specific case we benefit greatly
from the centimeter emission to refine our results. The best-fit model
using MIR data alone has $\chi^{2}_{\rm IR} = $ 0.04 and produces
parameters of $m_{\star} = 2\,M_{\sun}$, a
core of mass $M_{c} = 10\,M_{\sun}$ and $\Sigma_{\rm cl} =$ 3.2
g\,cm$^{-2}$ clump. However, we disfavor this model since from the
radio observations it is obvious that this source has a temperature
high enough to emit UV photons that has already formed an UC/HC HII
region. Therefore, our best matching model as based on the
$\chi^{2}_{IR}$ = 0.29 and $\chi^{2}_{all\_inner} =$ 4.52 has
$m_{\star} = 12\,M_{\sun}$, a core of mass $M_{c} = 50\,M_{\sun}$
and $\Sigma_{\rm cl} =$ 0.1 g\,cm$^{-2}$ clump. These parameters are also
more consistent with the ones reported by
\citet{2010MNRAS.402.2583K}. 

\emph{IRAS 07299-1651:}\, This source also has a relatively limited
amount of data in the FIR and in the centimeter to constrain the
models. The best-five ZT models using only the IR data indicate that a
protostar in the range of 8--16 $M_{\sun}$, in relatively
low-$\Sigma$ clumps and mainly in cores of more than a couple hundred solar
masses can fit well the observations. When including our centimeter
wavelength emission for this source, our best matching model has
$\chi^{2}_{\rm IR} = $ 0.90 and $\chi^{2}_{\rm all\_inner} =$ 3.46 with
parameters of $m_{\star} =\,12 M_{\sun}$, a core of mass $M_{c} =\,
240 M_{\sun}$ and a $\Sigma_{cl} =$ 0.1 g\,cm$^{-2}$ clump. A similar
fit is given by the second best model based on the
$\chi^{2}_{\rm all\_inner} =$ 3.54 with the only difference being the mass
of the core is slightly lower. 

\emph{G35.20$-$0.74:}\, The best-fit model using the IR SED alone also
corresponds to the best matching model when adding the centimeter data
emission with $\chi^{2}_{\rm IR} = $ 2.60 and $\chi^{2}_{\rm all\_inner} =$
19.35 with parameters of $m_{\star} =\,16 M_{\sun}$, a core of mass
$M_{c} =\,480 M_{\sun}$ and a $\Sigma_{cl} =$ 0.1 g\,cm$^{-2}$
clump. The source detected at the \emph{Inner} scale (or source 8a in
\citealt{2016A&A...593A..49B}) is slightly extended at 6 cm and it is blended with other components, however it is
a point source at 1.3 cm and 0.7 cm. Thus, it is possible then that we
are missing part of the flux at 6 cm and this may explain why the models
seem to overestimate the expected free-free emission at 6 cm. Our resulting protostellar mass of 16 $M_{\sun}$ is
consistent with the value of of 18 $M_{\sun}$ estimated by
\citet{2013A&A...552L..10S} when fitting the velocity field of the
core with a rotating Keplerian disk, although they argue that this
value corresponds to the total mass of a binary system in core B. Our
results disfavor the models with protostellar mass $m_{\star} =\,12
M_{\sun}$ and with relatively low core masses and high $\Sigma_{cl}$
clumps, since the expected free-free emission for these models is many
orders of magnitude lower than the observed one.  

\emph{G45.47$+$00.05:}\, The observed centimeter continuum emission
from G45.47$+$00.05S (which is the dominant source in the region) is
significantly higher (by a few orders of magnitude) than the predicted
free-free emission from any of our IR-derived ZT models. This is
despite the fact that the best two models from the ZT results are
already predicting a rather high-mass protostar in the range of 30--50
$M_{\sun}$ embedded in a massive core, a relatively high $\Sigma$
clump and low accretion rates in the range of $\sim$10$^{-5}
M_{\sun}$\,yr$^{-1}$. Among the eight sources presented in this study
this is the only case where none of the free-free models can describe
the centimeter emission observed in the source. This region is also
the most luminous and likely the most evolved one in this sample,
which leads us to think that the radio emission from this source may
be boosted by the process of photoevaporation. Photoevaporation is not
yet accounted for in the TTZ model, which only considered
photoionization of the magnetocentrifugally-driven wind. However, as
the protostar increases its mass above $\sim20\:M_\odot$, the ionizing
radiation becomes dramatically stronger, creating a photoevaporation
flow from the disk and infall envelope that is exposed by the outflow
cavity wall, i.e., a wind driven by ionized gas-pressure. By enhancing
the mass of ionized gas, the centimeter continuum emission is expected
to be much higher than that predicted by TTZ model without such a
photoevaporation flow.

\emph{IRAS 20126$+$4104:}\, Including the centimeter continuum fluxes to
better sample the SED, we can see in Figure \ref{fig:SEDs} that the
best five models seem to be overestimating the expected free-free
emission. Therefore, after exploring other resulting models from the
\citetalias{2017ApJ...843...33D} study and the ZT models, we found a
matching model with $\chi^{2}_{IR}$ = 3.38 and $\chi^{2}_{\rm
  all\_inner} = $ 5.42, which yields protostellar parameters of
$m_{\star} = 8\:M_{\sun}$, core mass $M_{c}= 240\:M_{\sun}$ and
$\Sigma_{\rm cl}=$ 0.1 g\,cm$^{-2}$. This protostellar mass is consistent
with the value of 7--10 $M_{\sun}$ estimated by
\citet{2005A&A...434.1039C,2011A&A...526A..66M,2014A&A...566A..73C}
from methyl cyanide emission that is likely tracing a Keplerian disk
around the protostar. However, with similar methods \citet{2016ApJ...823..125C}
estimated a protostellar mass of 12 $M_{\sun}$. Therefore either of
our best-two models presented in Table \ref{model_parameters} for this
source could be applicable.  

\emph{Cepheus A:}\, The best matching model for this source after
using the extended SED from centimeter to NIR emission has
$\chi^{2}_{\rm IR} = $ 2.43 and $\chi^{2}_{\rm all\_inner} =$ 7.13 and is
constrained to a protostellar mass of $m_{\star} =\ 12\: \,M_{\sun}$, a
relatively massive core of $M_{c}= 480\: \,M_{\sun}$ and a $\Sigma_{\rm cl}=$
0.1 g\,cm$^{-2}$ clump. Our result is consistent with the kinematic
masses estimated for the central source HW2 which are in the range of
10--20 $M_{\sun}$ \citep[e.g.,][]{2017A&A...603A..94S}.

\emph{NGC 7538 IRS9:}\, Our results using the extended SED favor all our
models with a protostellar mass of $m_{\star} = 16\: \,M_{\sun}$, with a
relatively massive core in the range of 320--480 $M_{\sun}$ and a
$\Sigma_{\rm cl}=$ 0.1 g\,cm$^{-2}$ clump.  
Based
on our results, we disfavor the model that has a 12 $M_{\sun}$
protostar, embedded in a core of $M_{c}= 60\: \,M_{\sun}$ and
$\Sigma_{\rm cl}=$ 3.2 g\,cm$^{-2}$, since the predicted free-free emission
from such model is significantly lower than the observed one.


\begin{deluxetable}{l c c c c c c c c c c c c c c}
\tabletypesize{\scriptsize}
\tablecaption{Parameters of the Best-fitted Models of Zhang \& Tan and Tanaka et al. 2016  \label{model_parameters}}
\tablewidth{0pt}
\tablewidth{0pt}
\tablehead{
\colhead{Region}                  & 
\colhead{$\chi^{2}_{all\_inner}$}   &
\colhead{$\chi^{2}_{all\_inter}$}  &
\colhead{$\chi^{2}_{all\_SOMA}$}&
\colhead{$\chi^{2}_{\rm IR}$}  &
\colhead{$M_{c}$}   &
\colhead{$\Sigma_{\rm cl}$ }               & 
\colhead{$R_{c}$}  & 
\colhead{$m_{\star}$}             & 
\colhead{$\theta_{\rm view}$} & 
\colhead{$A_{V}$}        &
\colhead{$M_{\rm env}$}  &
\colhead{$\theta_{\rm w,esc}$}  &
\colhead{$\dot{M}_{\rm disk}$} & 
\colhead{$L_{bol}$} 
 \\
\colhead{}                            & 
\colhead{}                      & 
\colhead{}                      & 
\colhead{}                      & 
\colhead{}                      & 
\colhead{($M_{\sun}$)}                      & 
\colhead{(g\,cm$^{-2}$)}                      & 
\colhead{(pc)($^{\prime\prime}$)}                      & 
\colhead{($M_{\sun}$)}                      &  
\colhead{($^{\circ}$)}                      & 
\colhead{(mag)}                      &
\colhead{($M_{\sun}$)}                      &
\colhead{($^{\circ}$)}                      &
\colhead{($10^{-4}M_{\sun}$\,yr$^{-1}$)}                      &
\colhead{($10^{4}L_{\sun}$)}                              \\[-20pt]\\}
\startdata
AFGL 4029&	 11.07&   1.71&	  1.62   &	1.74&	80&	0.1&	0.21(22)&	8&	74&	0.0&	62&	27&	0.5&	0.97 \\	
         &       36.87&   9.03&   1.51   &       1.36&   80&	0.1&	0.21(21)&	12&	88&	0.0&	47&	40&	0.5&	1.6 \\
         &       43.55&   12.31&  1.52   &       1.16&  160&	0.3&	0.17(17)&	48&	88&	15.2&	14&	77&	1.1&	0.34 \\	
         &       51.95&   16.83&  1.68   &       1.07&	30&	1.0&	0.04(4)&	12&	62&	13.1&	6&	53&	1.9&	4.1 \\
         &       58.35&   18.29&  2.16   &       1.55&   50&	0.3&	0.09(10)&	12&	51&	15.2&	22&	46&	1.0&	2.4 \\	
         &       60.05&   18.02&  2.08   &       1.47&   60&	0.3&	0.10(10)&	16&	62&	4.0&	19&	56&	1.1&	3.6 \\[9pt]
AFGL 437 &      4.52  & \nodata& 0.42   &        0.29&	50&	0.1&	0.16(17)&	12&	80&	7.5&	15&	59&	0.3&	1.4 \\	
         &      17.89 & \nodata& 1.66   &        0.32&   50&	0.3&	0.09(9)&	16&	77&	2.6&	 8&	68&	0.7&	3.1 \\
         &      49.37 & \nodata& 5.28   &        0.53&   60&	1.0&	0.06(6)&	24&	89&	16.8&	 5&	71&	1.9&	9.3 \\
         &      334.54& \nodata& 36.74  &        0.51&   20&	3.2&	0.02(2)&	 4&	39&	0.0&	12&	34&	3.1&	0.3 \\
         &	399.67& \nodata& 42.59  &	0.04&	10&	3.2&	0.01(1)&	 2&	39&	0.0&	 6&	35&	1.8&	0.3 \\[9pt]		
IRAS 07299&      3.46& \nodata&  1.47   &       0.90&  240&	0.1&	0.36(44)&	12&	89&	48.5&	211&	19&	0.9&	2.0 \\
         &       3.54& \nodata&  1.59   &       1.07&  200&	0.1&	0.33(40)&	12&	89&	51.5&	174&	20&	0.8&	2.0 \\
         &       22.65& \nodata&  16.42   &	0.62&  240&	0.1&	0.36(44)&	8&	89&	6.1&	226&	13&	0.7&	1.1 \\
         &       26.21& \nodata&  13.07   &       0.86&	80&	0.3&	0.12(14)&	16&	89&	12.1&	42&	42&	1.5&	4.2 \\
         &       70.69& \nodata&  46.46  &       1.16&  400&	0.1&	0.47(57)&	8&	62&	0.0&	386&	10&	0.8&	1.0 \\[9pt]			
G35.20$-$0.74&   19.35&    34.03& 21.00   &	2.60&  480&	0.1&	0.51(48)&	16&	48&	40.4&	440&	15&	1.2&	3.9 \\	
         &       26.10&    19.89&  13.05   &          2.76&  200&	0.3&	0.19(17)&	12&	22&	43.4&	173&	17&	1.9&	4.0 \\
         &       33.14&    15.25&  9.85   &          2.79&   80&	3.2&	0.04(3)&	12&	39&	15.2&	58&	22&	8.4&	5.0 \\	
         &      135.79&   472.57&  328.48  &          2.62&  100&	3.2&	0.04(4)&	12&	34&	28.3&	77&	20&	9.4&	5.2 \\	
         &      158.33&  521.58&  356.98  &          2.78&  320&	0.1&	0.42(39)&	24&	68&	81.8&	256&	27&	1.2&	8.4 \\[9pt]
G45.47$+$0.05&  40.94& \nodata&	 9.53   &	1.20&  200&	3.2&	0.06(1)&	32&	89&	61.6&	140&	25&	16.9&	46.0 \\		
         &      57.05& \nodata&  12.91   &          1.33&  320&	1.0&	0.13(3)&	48&	89&	46.5&	200&	35&	9.3&	50.9 \\
         &      68.80& \nodata&  15.74   &          1.68&  320&	1.0&	0.13(3)&	32&	68&	15.2&	252&	24&	8.2&	27.4 \\	
         &      71.24& \nodata&  16.22   &          1.62&  240&	1.0&	0.11(3)&	32&	83&	2.0&	170&	30&	7.2&	25.7 \\
         &      79.60& \nodata&  18.14   &          1.71&  240&	1.0&	0.11(3)&	24&	55&	0.0&	192&	23&	6.6&	17.2 \\[9pt]
IRAS 20126&     5.42 & \nodata&	 2.99   &	3.38&  240&	0.1&	0.36(45)&	8&	71&	24.2&	226&	13&	0.7&	1.0 \\
         &      53.82 & \nodata&  2.48   &          2.38&  80&	0.3&	0.12(15)&	12&	44&	73.7&	53&	31&	1.4&	3.4 \\	
         &      106.35& \nodata&  3.44   &          2.29&  200&	0.1&	0.33(41)&	12&	89&	65.7&	174&	20&	0.8&	2.0 \\
         &      110.91& \nodata& 3.21   &	1.88&  80&	0.3&	0.12(15)&	16&	80&	33.3&	42&	42&	1.5&	4.2 \\		
         &      117.92& \nodata&  3.70   &          2.40&  100&	0.3&	0.13(16)&	16&	51&	68.7&	61&	36&	1.6&	4.5 \\
         &      136.40& \nodata&  3.86   &          2.16&  120&	0.3&	0.14(18)&	24&	74&	69.7&	57&	47&	1.8&	9.3 \\[9pt]	
Cepheus A&      7.13&  	  9.18&  \nodata&       2.43&  480&	0.1&	0.51(150)&	12&	89&	84.8&	457&	12&	1.1&	2.4 \\
         &      50.64&  72.00&  \nodata&       2.31&  160&	0.3&	0.17(49)&	12&	29&	100.0&	135&	20&	1.8&	3.8 \\	
         &      57.36&  89.49&  \nodata&       3.06&  120&	0.3&	0.14(42)&	12&	65&	62.6&	93&	24&	1.6&	3.6 \\
         &      63.71&  93.12&	\nodata&    2.23&  160&	0.3&	0.17(49)&	16&	44&	95.9&	125&	26&	2.0&	5.0 \\		         
         &       85.10&   126.97& \nodata&       2.78&  160&	0.3&	0.17(49)&	24&	83&	100.0&	98&	37&	2.2&	9.9 \\[9pt]
NGC 7538 &      13.18&  \nodata&  2.83   &         0.19&  320&	0.1&	0.42(32)&	16&	39&	2.02&	281&	19&	1.1&	3.7 \\
         &      14.32&  \nodata& 3.09 &   0.15&  400&	0.1&	0.47(36)&	16&	22&	22.2&	364&	17&	1.1&	3.8 \\		
         &      14.44&  \nodata&  3.52   &         0.42&  480&	0.1&	0.51(40)&	16&	22&	18.2&	440&	15&	1.2&	3.8 \\         	
         &      31.35&  \nodata&  1.22   &         0.35&  240&	0.1&	0.36(28)&	24&	39&	52.5&	171&	33&	1.1&	8.2 \\
         &      149.39& \nodata&  66.64  &         0.53&  60&	3.2&	0.03(2)&	12&	34&	21.2&	38&	27&	7.6&	5.0 \\[9pt]

\enddata
\tablecomments{\footnotesize{Models are listed from best to worst as measured
by the reduced $\chi^{2}$ from the inner scale
($\chi^{2}_{all\_inner}$) of each region. Parameters $\chi^{2}_{\rm IR}$,  $\theta_{\rm view}$ and $A_{V}$ from the best five models reported in \citetalias{2017ApJ...843...33D} have been updated reflecting upgrades in the ZT model. First model for sources AFGL 4029 and IRAS 20126$+$4104 are new estimates corresponding to better fits (see \S\ref{sec:models}).}}
\end{deluxetable}

\begin{figure}[htbp]
\centering
\begin{tabular}{cc}
\hspace*{\fill}%
\includegraphics[width=0.48\textwidth,  trim = 20 20 20 15, clip, angle = 0]{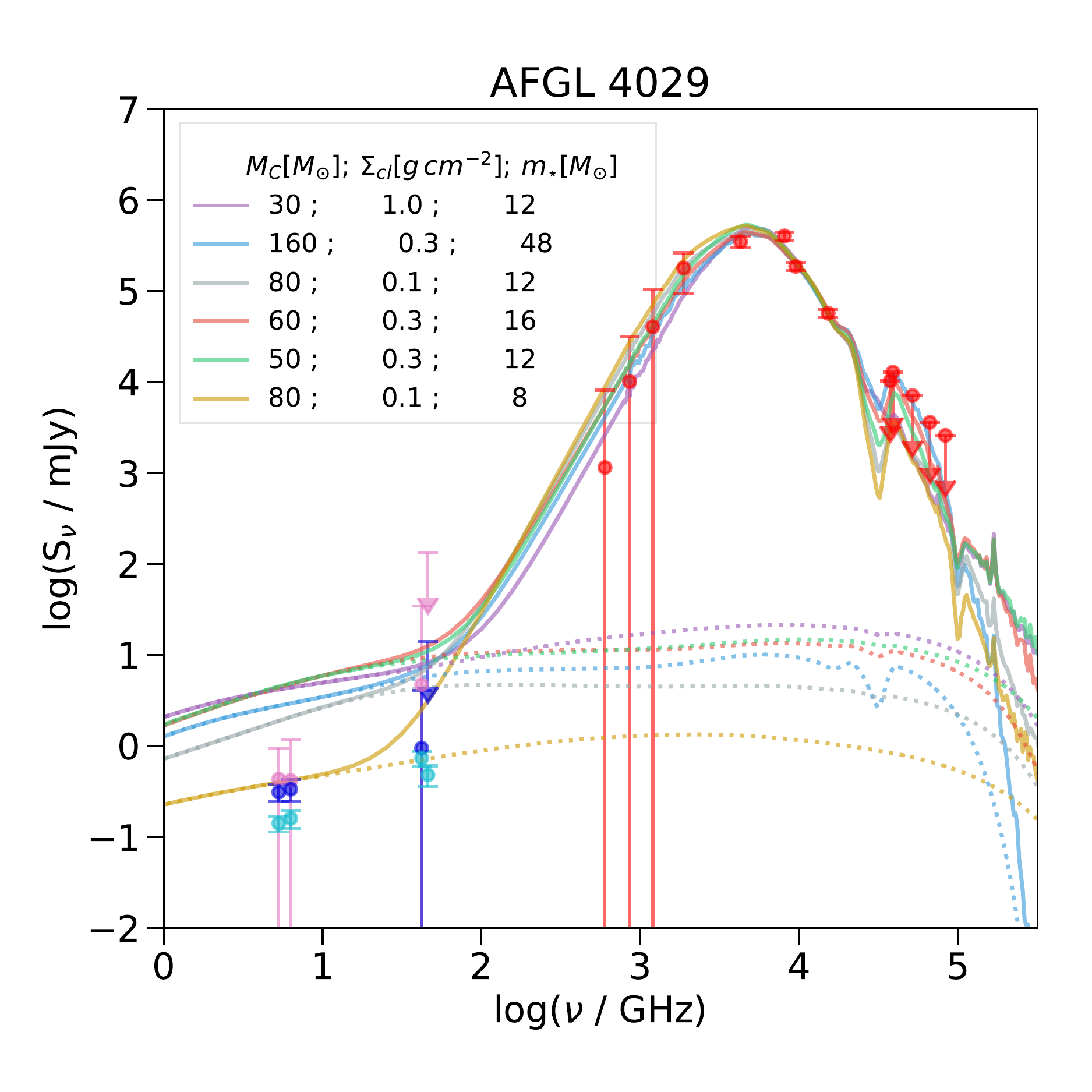}\hspace{-0.1cm} &
\includegraphics[width=0.48\textwidth,  trim = 20 20 20 15, clip, angle = 0]{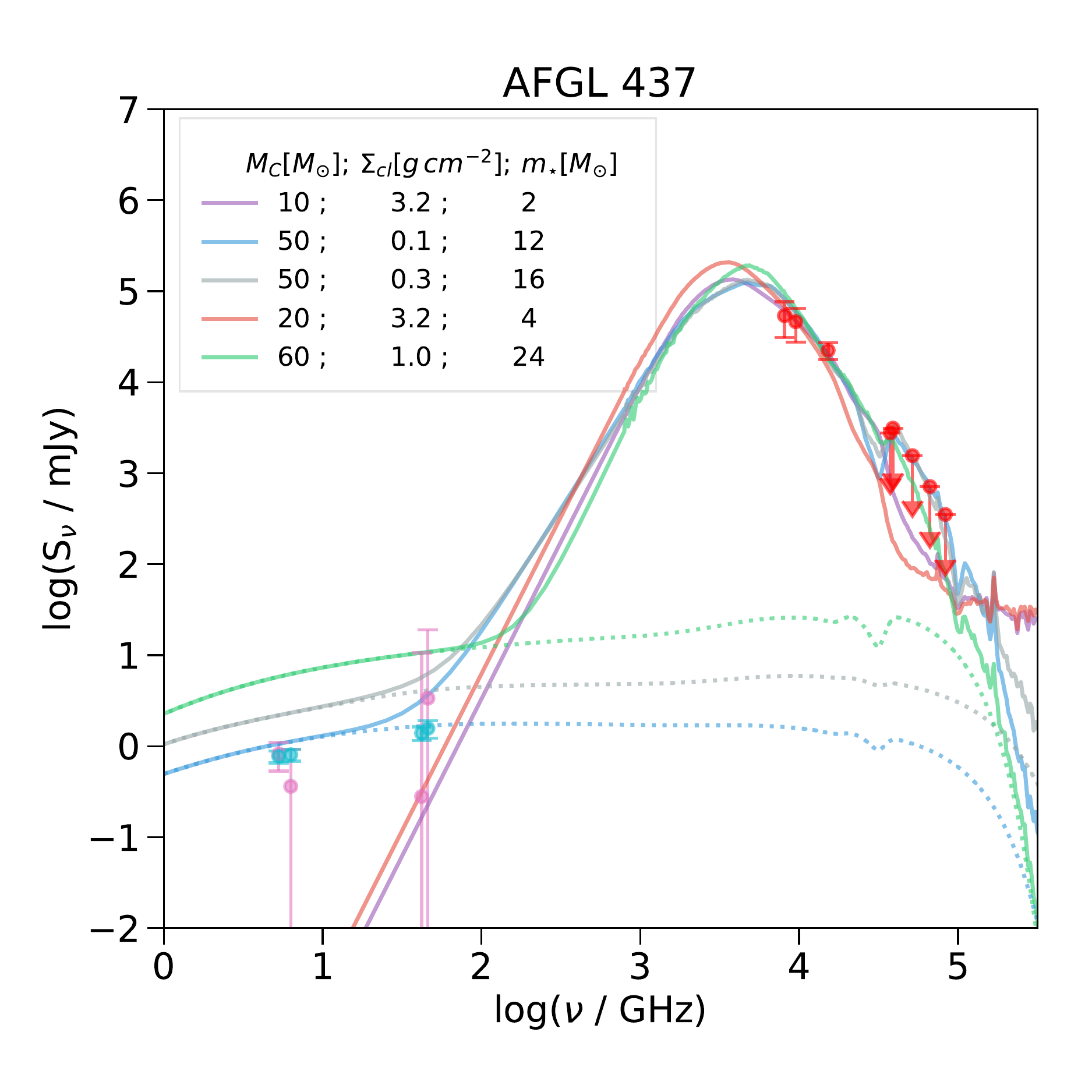}\\
\includegraphics[width=0.48\textwidth,  trim = 20 20 20 15, clip, angle = 0]{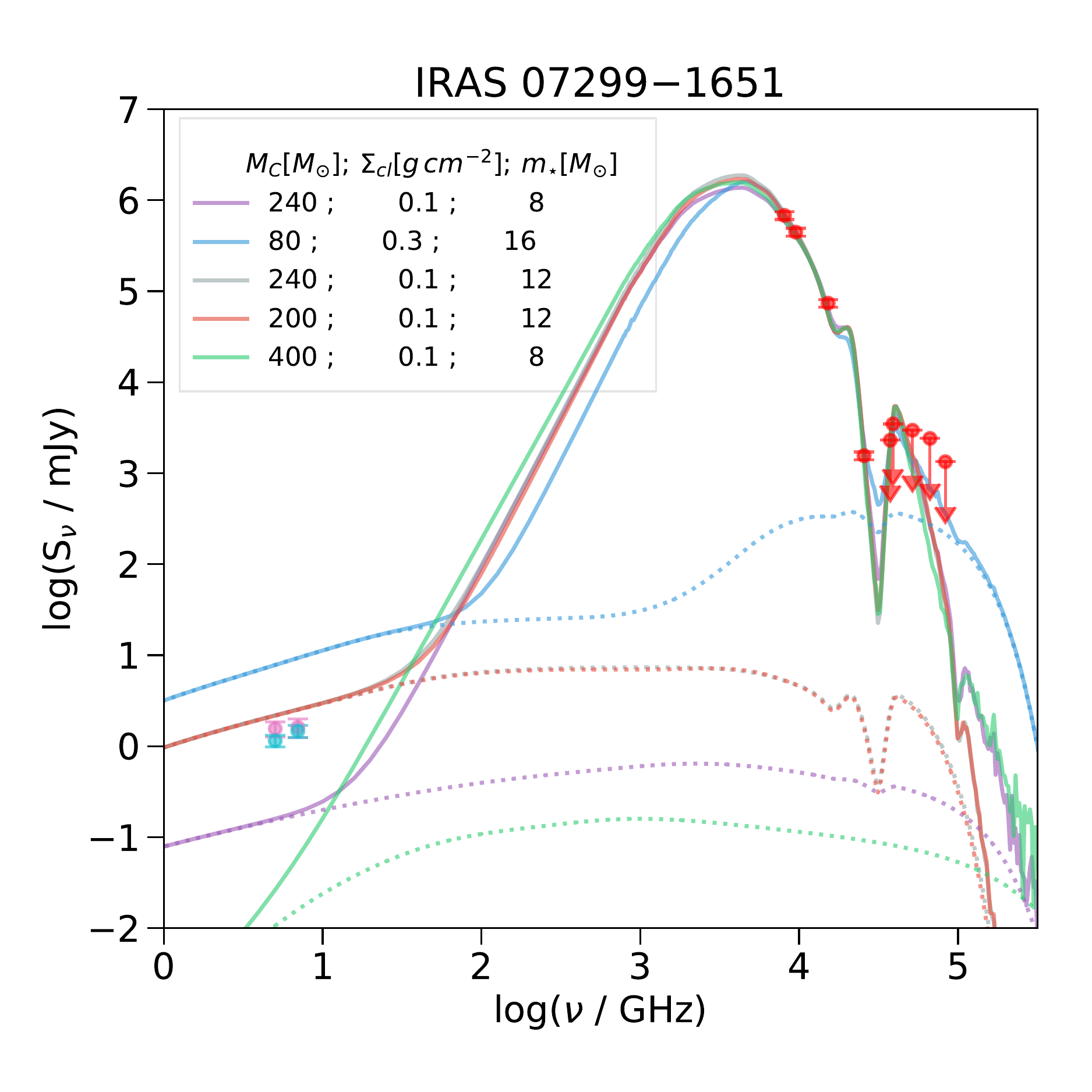}\hspace{-0.1cm} &
\includegraphics[width=0.48\textwidth,  trim = 20 20 20 15, clip, angle = 0]{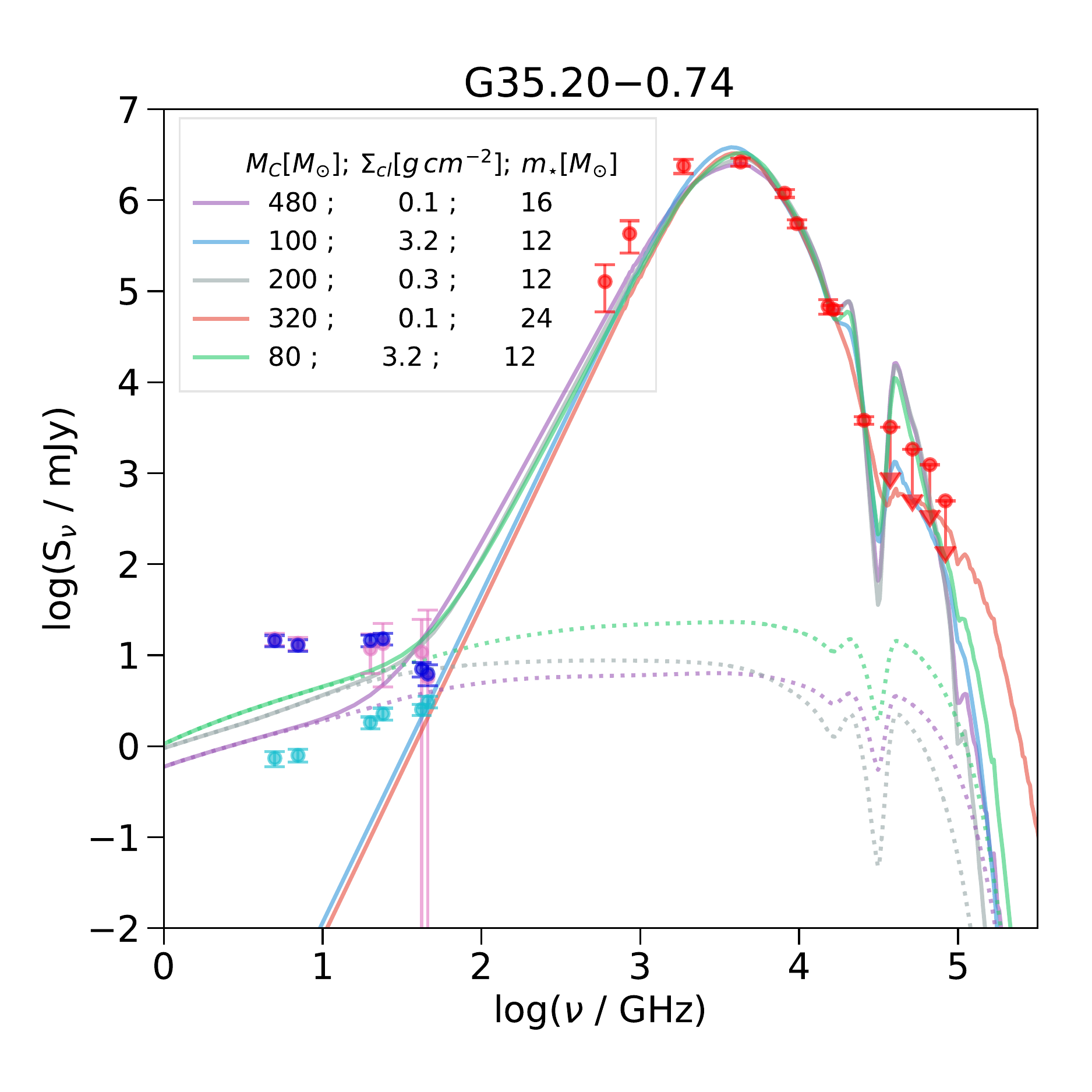}\\

\end{tabular}
\caption{\small{
Comprehensive SEDs and models of the protostars.  Red circles show IR
data for the SOMA apertures as measured by
\citetalias{2017ApJ...843...33D}.  Solid colored lines are 
best fits to the IR data (except for yellow lines) from the
\citetalias{2018ApJ...853...18Z} models, which have been augmented by including free-free emission
from the \citetalias{2016ApJ...818...52T} model. The dotted colored
lines correspond to the estimated free-free component using
\citetalias{2016ApJ...818...52T}, which only exceeds the thermal dust
component at long wavelengths. For sources AFGL 4029 and IRAS
20126$+$4104 the yellow solid and dotted lines correspond to the best
new model estimate of the total emission and the free-free emission,
respectively. The colored circles correspond to the flux density as a
function of frequency for each scale (magenta: SOMA; blue:
Intermediate; cyan: Inner). Error bars are explained in
\S\ref{spectral_indices}.  }}
 \label{fig:SEDs}
\end{figure}

\begin{figure}[htbp]
\centering
\begin{tabular}{cc}
  \ContinuedFloat
\hspace*{\fill}%
\includegraphics[width=0.48\textwidth,  trim = 20 20 20 15, clip, angle = 0]{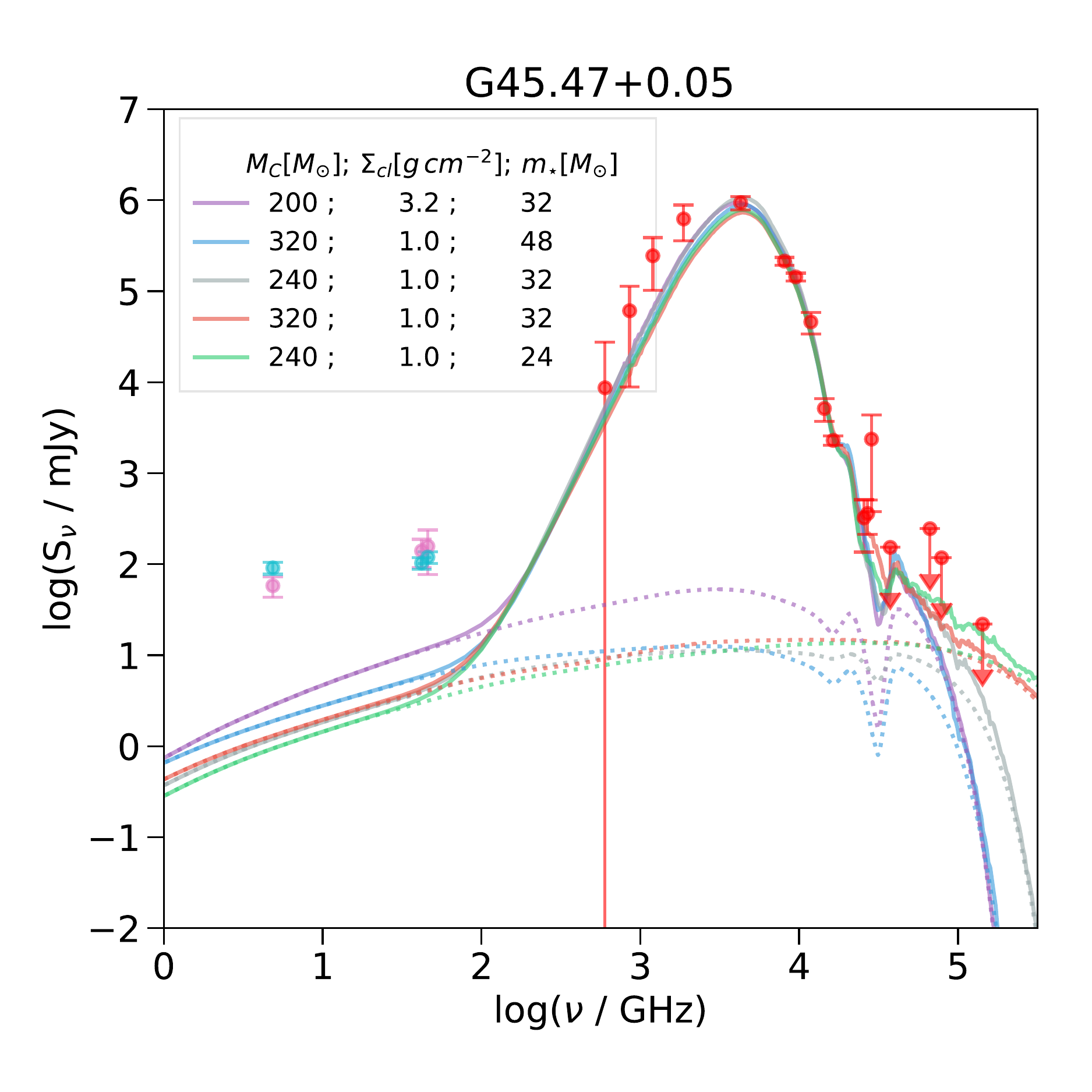}\hspace{-0.1cm} &
\includegraphics[width=0.48\textwidth,  trim = 20 20 20 15, clip, angle = 0]{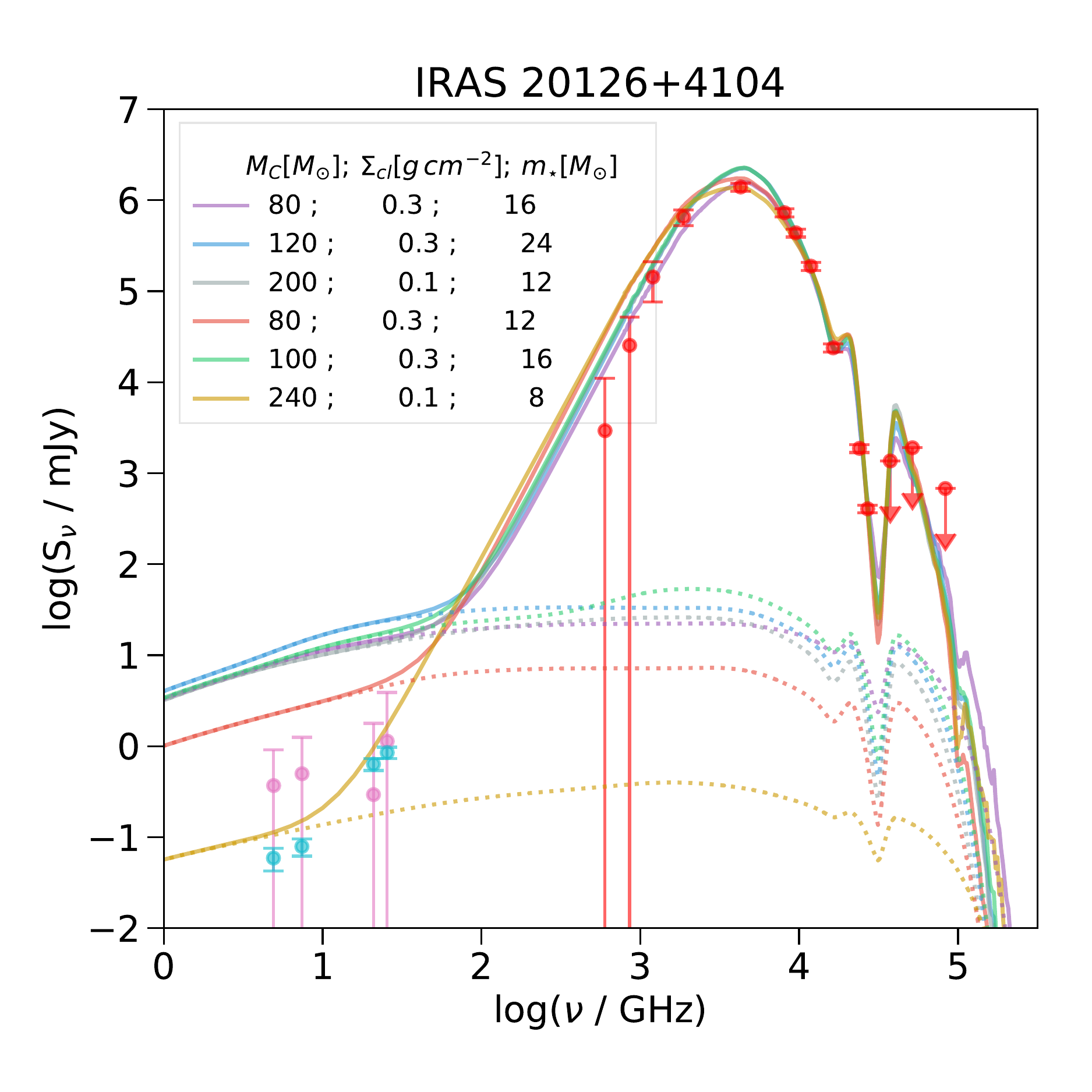}\\
\includegraphics[width=0.48\textwidth,  trim = 20 20 20 15,  clip, angle = 0]{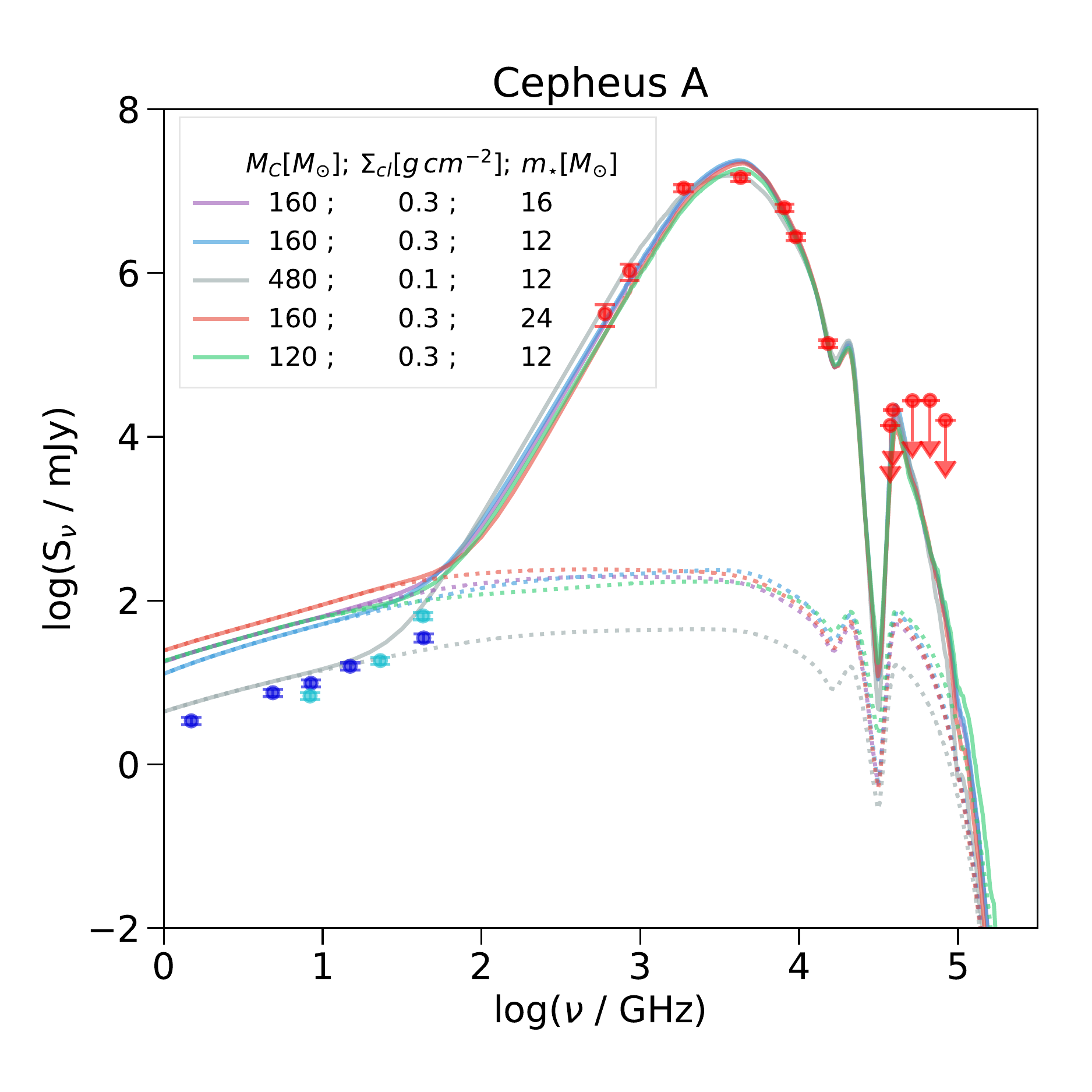}\hspace{-0.1cm} &
\includegraphics[width=0.48\textwidth,  trim = 20 20 20 15, clip, angle = 0]{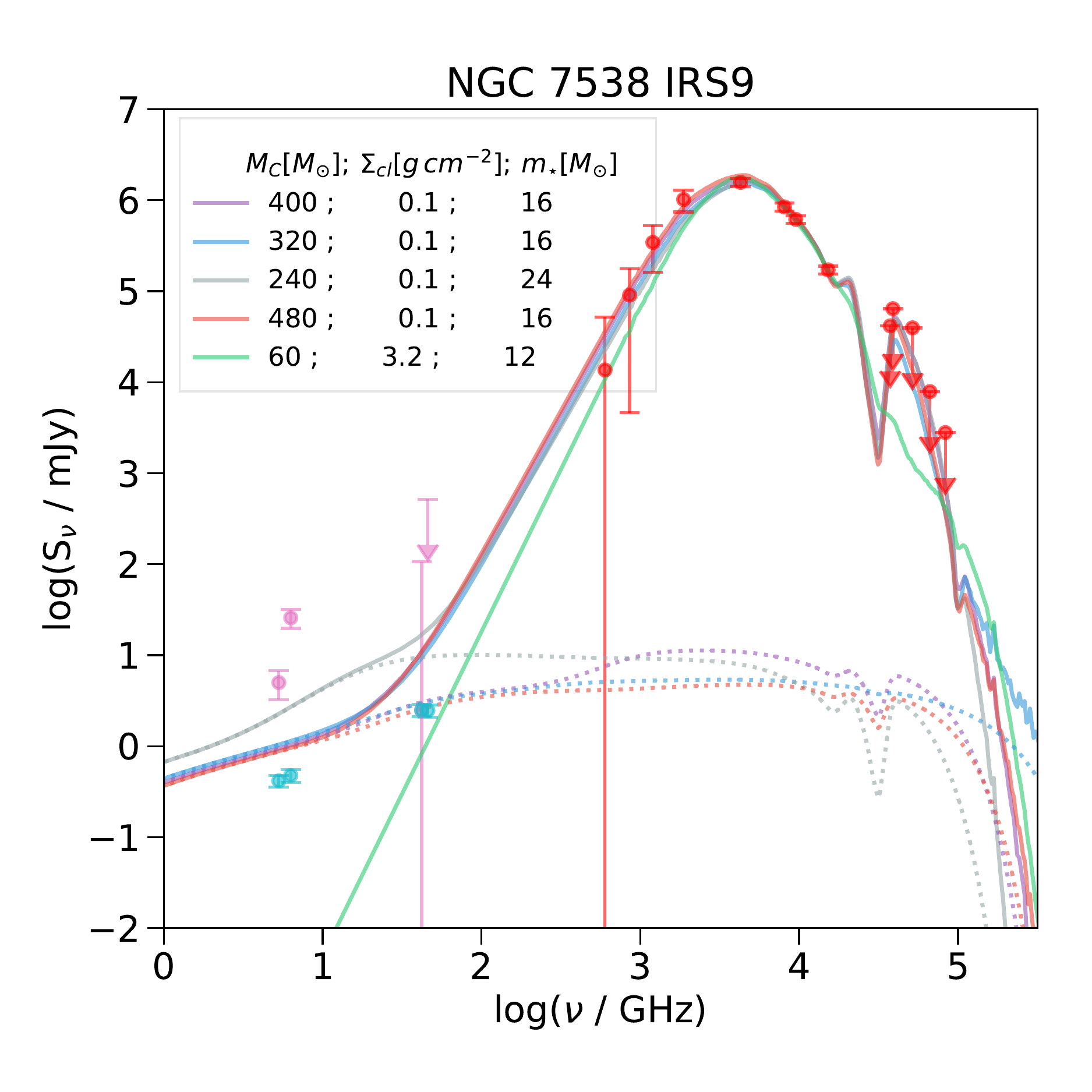}\\
\end{tabular}

 \hspace*{\fill}%
\caption{\small{Continued.}}
 \label{fig:SEDs}
\end{figure}

\section{Discussion}\label{discussion}

\citetalias{2017ApJ...843...33D} tested the ZT radiative transfer
model by fitting MIR and FIR data on the same sample that we present
in this paper. By obtaining good fits to the data, they showed that
high-mass protostar models based on Core Accretion that are physically
self-consistent and scaled-up from those developed for lower-mass
protostars can be a reasonable description of the sources. However,
these solutions from simple SEDs are not unique and yield a range of
values for the main parameters that need to be further
constrained. Centimeter continuum emission is expected even for early
stages in the formation of high-mass stars (see
\citealt{2016ApJS..227...25R}) and we have utilized it here as an
important extra diagnostic and test of the protostellar models, i.e.,
of their ionizing luminosity. Using the initial range of parameters
that resulted from the ZT model we were able to use the
\citetalias{2016ApJ...818...52T} photoionization model to predict the
free-free emission expected for these initial conditions. We found
that extended SEDs that include longer wavelengths in the centimeter
regime help to break the degeneracies of the main physical parameters
such as the mass of the core, the mass surface density of the clump
and especially the mass of the protostar. Furthermore, our results are
consistent with values estimated from others methods like dynamical
protostellar masses.

For the subsample of Type II SOMA sources presented in this paper and
that are associated with outflowing material, we generally found
protostars in the range of $\sim$8--24 $M_{\sun}$, except for
G45.47$+$0.05. The centimeter wavelength emission detected towards
G45.47$+$0.05 is brighter than for the rest of the sample and none of
the predicted free-free emission models fit these long wavelength
observations. We speculate that the origin of the detected radio
emission is due to an extra contribution from photoevaporation of the
disk and infall envelope. However further analysis and modeling is
required to fully understand its nature. For instance, an
observational diagnostic to differentiate between a MHD disk wind and
a photoevaporation flow is through the width of hydrogen recombination
lines (HRLs). A disk wind driven by magnetocentrifugal forces provides
a broader width of $>100{\rm\:km\:s^{-1}}$
\citep{2011ApJ...732L..27J}, while a photoevaporation flow has a
narrower profile of $<100{\rm\:km\:s^{-1}}$
\citep{2014ApJ...796..117G}. From the modeling side, we defer the
analysis of more evolved sources that require the addition of
photoevaporation components in the \citetalias{2016ApJ...818...52T}
for a later study. We also note that contributions from shock
ionization, which may be especially relevant for the relatively weak
centimeter emission in extended jet knots, is not yet included in the
models.

\begin{figure}[htbp]
\centering
\includegraphics[width=1.0\textwidth,   clip, angle = 0]{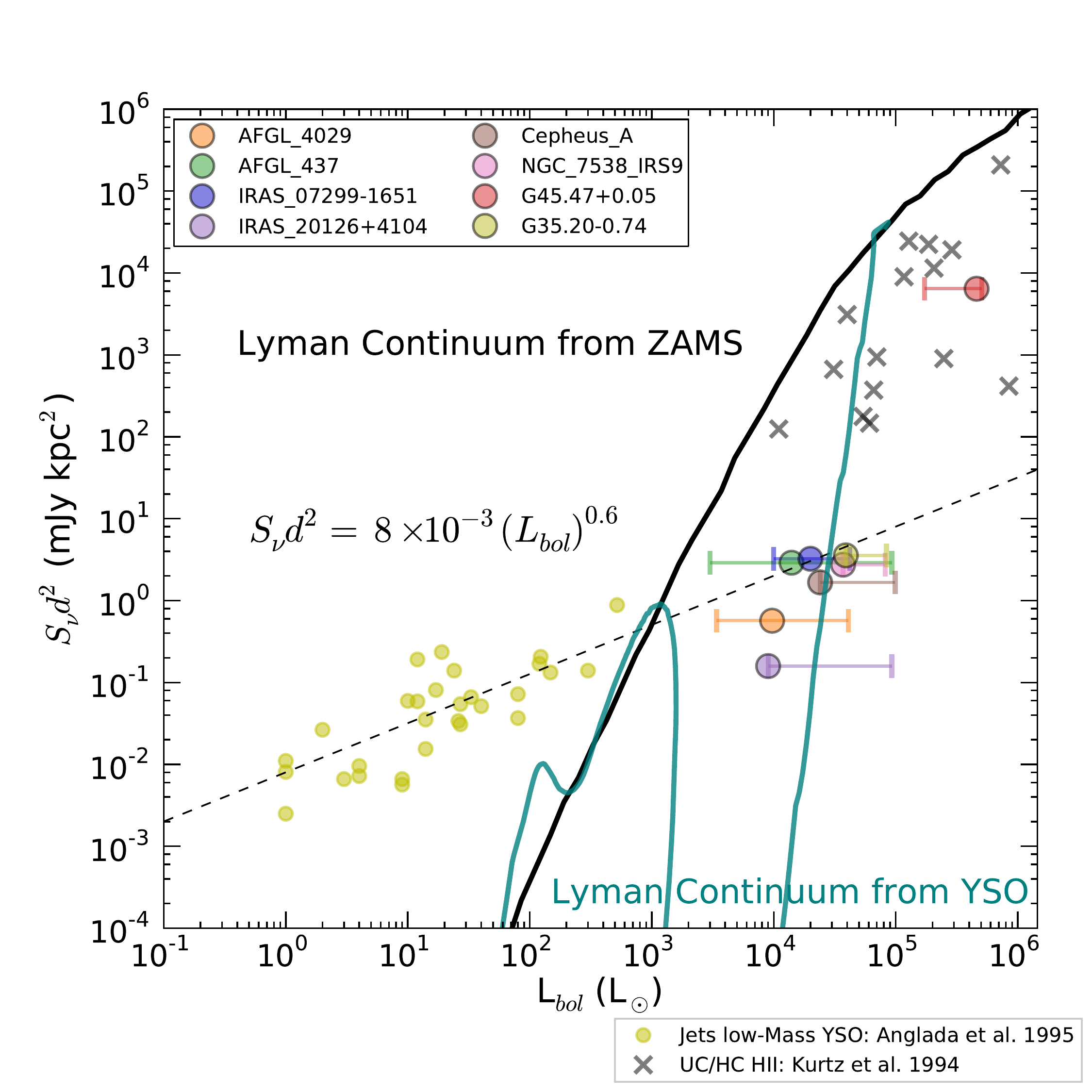}
\caption{\small{
Radio luminosity at 5 GHz for the \emph{Inner} scale as a function of
the bolometric luminosity of our eight SOMA sources. The bolometric
luminosity is given by our results of the best model (lower $\chi^{2}$
for the \emph{Inner} scale) and the error bar corresponds to the range
of bolometric luminosities of the best models for each source listed
in Table \ref{model_parameters}. The yellow circles represent ionized
jets toward low-mass stars from \citet{1995RMxAC...1...67A}. The dashed line
shows a power law relation for these sources,
given by \citet{2015aska.confE.121A}: $S_{\nu}d^{2} = 8\times10^{3}
(L_{bol})^{0.6}$.  The $\times$ symbols are UC and HC HII regions from
\citet{1994ApJS...91..659K}. The black and the cyan continuous lines
are the radio emission expected from optically thin HII regions
powered by a ZAMS star \citep{1984ApJ...283..165T} and from a YSO
(\citetalias{2016ApJ...818...52T}), respectively. Note, these models
assume all of the ionizing photons are reprocessed by the HII region,
i.e., with zero escape fraction.}}
 \label{fig:Anglada_plot}
\end{figure}

As an additional diagnostic to understand the nature of our sources,
in Figure~\ref{fig:Anglada_plot} we compare the bolometric luminosity
with the radio luminosity at 5 GHz from the \emph{Inner} scale of our
eight SOMA sources (the fluxes have been scaled to $\nu =$~5~GHz using
the spectral index estimated at that scale). The bolometric luminosity
is given by our results of the best model (lowest $\chi^{2}$ for the
\emph{Inner} scale) and the error bar corresponds to the range of
bolometric luminosities from the models listed in Table
\ref{model_parameters}. Additionally, we show as yellow circles the
radio luminosity from lower-mass protostars associated with ionized
jets from \citet{1995RMxAC...1...67A}. We scaled their fluxes from 3.6
cm using a factor of 0.74 assuming that those sources have a spectral
index $\alpha=0.6$, which is the canonical value of ionized jets.
A power law fit to these data of $S_{\nu}d^{2} = 8\times10^{3}
(L_{bol})^{0.6}$ is shown with a dashed line. We also show several
UC/HC HII regions from \citet{1994ApJS...91..659K} represented with
$\color{gray} \times $ symbol. The continuous black line is the radio emission from an optically thin
HII region given the expected Lyman continuum luminosity of a single
zero-age main-sequence (ZAMS) star at a given luminosity
\citep{1984ApJ...283..165T}. The cyan continuous line corresponds to
the expected radio emission that arises from photoionization from a
protostar as predicted by the \citetalias{2016ApJ...818...52T} model,
also for optically thin conditions at 5~GHz. This specific
evolutionary stellar model corresponds to the fiducial case which
starts with a core mass of $M_{c}= 60\:M_{\odot}$ and a mass surface
density of ambient clump of $\Sigma_{\rm cl}=$~1~g~cm$^{-2}$.

From Figure \ref{fig:Anglada_plot} we see that the ionized material
towards  several of our sources (except G45.47$+$0.05) appear to
follow the same power law relation found by
\citet{2015aska.confE.121A}, which may indicate that a universal
mechanism based on shock ionization (see \citealt{2018A&ARv..26....3A}
for a review) is still relevant for these sources. However, they also
match quite well with the example model (cyan line) from
\citetalias{2016ApJ...818...52T}, which is based on
photoionization. One must bear in mind that there is a very large
dynamic range present for the radio luminosities. A more comprehensive
theoretical model that includes both shock and photoionization may be
needed to better model these sources. Still, at the highest bolometric
luminosities, i.e., as sampled by G45.47$+$0.05, is seems likely that
the sources are in a photoionization dominated regime, and a
photoionized outflow model may be relevant to many HC HII region
sources.
We note that in our detailed modeling of radio SEDs as applied to the
sources, there is typically a high escape fraction of ionizing photons
from the source, as well as loss of ionizing photons by absorption by
dust. Thus these models are generally lower in their radio flux than
the simple extrapolation shown in Fig.~\ref{fig:Anglada_plot} by the
cyan line. As we have mentioned earlier, our detailed models likely
need an additional photoevaporative flow component to be able to
explain the strong radio fluxes of the source of G45.47$+$0.05.

Our main interest for this paper has been to measure the flux density
of the centimeter continuum sources associated to our regions and
compare them with the predicted free-free emission from the
\citetalias{2016ApJ...818...52T} model using the initial parameters
from the best-fitting results of the ZT model presented in
\citetalias{2017ApJ...843...33D}. Additionally, for completeness we
measured the flux densities of the detected radio sources at three
defined scales: the \emph{SOMA} scale that has the same size radius
used for the IR photometry for each region, the \emph{Intermediate}
scale that measures the flux density of radio detections that appear
aligned and that may be part of a radio jet and the \emph{Inner}
scale, which is the most localized region around the central
protostar. This approach helps us understand if the central source is
the more dominant one in the SOMA scale or if there are perhaps other
protostellar sources adding to the centimeter continuum emission as we
think is the case of G35.20$-$0.74 and possibly NGC 7538 IRS9 and IRAS
20126$+$4104.

We also attempted to understand the nature of the detected radio
continuum sources, but due to the difference in resolution of our data
at the given frequencies (except for regions G35.20$-$0.74 and IRAS
20126$+$4104) we are not able to have reliable estimates of the
spectral indices at the different scales studied in this
paper. However, if most of the detected radio sources associated with
our \emph{Inner} scale have a jet nature we will expect them to be
more compact at Q-band because ionized jets have a gradient of density
and are partially optically thick. Thus we expect the base of the jet
to be smaller at higher frequencies.
In order to determine reliable spectral indices, we require data with
similar resolutions that can be sensitive to the same scales.

Another interesting aspect that we can learn from these data is about
multiplicity. For all regions except AFGL 437, at least two sources
are detected within the \emph{SOMA} scale at the lower frequencies
(i.e., 6 cm data). Several of these detections appear elongated in the
same direction as the associated molecular outflow, e.g., in the case
of AFGL 4029, G35.20$-$0.74, IRAS 20126$+$4104 and Cepheus A, so it is
very likely that the central radio source at the \emph{Inner} scale
corresponds to the base of the ionized jet and the aligned radio
sources (if any) correspond to knots of the jet. This could also be
the case of IRAS 07299$-$1651, although at the moment we cannot ruled
out that the extended emission in our image for this region is due to
calibration errors.
Thus, for the above sources there is no strong evidence for stellar
multiplicity.
On somewhat larger scales in some sources we  find evidence for
other stellar sources.
Together with information from the literature, our presented data
reveal the detection of several variable radio sources within the
\emph{SOMA} scale, e.g., in AFGL 4029 and NGC 7538 IRS9. Furthermore,
IRAS 20126$+$4104 has at least two of these variable radio sources
that have spectral indices that are consistent with non-thermal
emission and are surrounding the high-mass protostar located at the
center of the core. One possible scenario for these radio variable
sources, at least for the one located at distances $<$2 kpc, is that
they correspond to flaring T-Tauri stars, indicating the presence of a
few lower-mass YSOs in the vicinity, at least in projection, of the
high-mass protostar. 

 The detectability (i.e., at 5$\sigma$ signal to noise ratio) of low-mass protostars in this sample is analyzed using the results of the Gould Belt survey, which is a large sample of low-mass YSOs observed with the VLA at 4.5 and 7.5 GHz \citep{2013ApJ...775...63D, 2014ApJ...790...49K, 2015ApJ...805....9O, 2015ApJ...801...91D, 2016ApJ...818..116P}. Of the bright low-mass protostars  detected in the Gould Belt survey, the two brightest ones are class III YSOs located in the Ophiuchus region with S$_{7.5GHz} =$ 8.51 mJy and 7.1 mJy \citep{2013ApJ...775...63D}. We find that such objects could be detected in our combined C-band images at level of  $\sim 5\sigma$, based on an average distance of $\sim1.8$ kpc and average image rms of  $6\,\mu$Jy/beam, but would not be detectable in our images at higher frequencies. This analysis applies to 5 regions (i.e., AFGL 4029, AFGL 437, IRAS 07299-1651, G35.20-0.74, IRAS 20126+4104), for which the C-band sensitivities and distances are very similar (to about $\pm 20\%$). We expect to be sensitive to such objects everywhere within the SOMA scale for these 5 regions since the noise in the primary beam corrected C-band image is essentially constant ($\sim$ 2$\%$). Also, these images have no bright sources causing sidelobes within the SOMA scale, thus we think that dynamic range is not an issue in the detection of low-mass protostars at the levels presented above.  For the other 3 regions we either do not have data at C-band or the images have lower sensitivity and therefore a detection of such objects is not expected, although we cannot rule out the presence of  T-Tauri stars brighter than those detected in the Gould Belt survey.    
Deeper VLA observations are needed for all the SOMA  regions to place more stringent constraints on the low-mass YSO population, but with the current observations there is no evidence for rich clusters of such YSOs around the high-mass protostars.

\section{Summary}\label{summary}

We have presented a pilot study mainly using public archival
interferometry data from the VLA to build extended SEDs from
centimeter emission to FIR to test theoretical models of high-mass star
formation forming via Core Accretion, in particular the
\citetalias{2016ApJ...818...52T} and \citetalias{2018ApJ...853...18Z} models. The
\citetalias{2016ApJ...818...52T} model  reproduces the SEDs of the IR and radio data for early-type
of sources before the existence of a strongly photoevaporated flow
contribution, which is not yet part of the model.  Our results
indicate that centimeter continuum emission is effective at breaking
degeneracies encountered in the IR-only analysis of the main physical
parameters such as the mass of the core, the mass surface density of
the clump and the mass of the protostar, with the main diagnostic
power coming from the strong dependence of ionizing luminosity of the
protostars as a function of the protostellar mass (though the
accretion rate also influences this given its effect on protostellar
evolution). Moreover, these resulting estimates of protostellar masses
appear more consistent with values obtained from other independent methods
such as dynamical mass estimates.

We also probed the presence of stellar multiplicity that is expected
to vary between Core Accretion and Competitive Accretion models of
high-mass star formation. We do not see large numbers of radio sources
that are likely to be other protostars or young stars around the
primary target, although a few lower-mass sources, perhaps variable
T-Tauri stars, are seen on larger scales around some of the high-mass
protostars. Of course, for these distant regions most low-mass
protostars may be too faint to see in the cm continuum, so deeper
observations are needed to better explore stellar multiplicity around
these high-mass protostars.

Also, in order to make a more uniform and systematic study of the
sources, specifically to understand the nature of the centimeter
wavelength emission associated to the SOMA survey regions and
interpret the centimeter continuum using the
\citetalias{2016ApJ...818...52T} models it will be ideal to have
similar resolutions and to be sensitive to the similar
scales. Expanding the sample size beyond the eight sources presented
here is also a high priority.

\acknowledgments  J.C.T. acknowledges support from NSF grant
AST 1411527 and several USRA/SOFIA grants in support of the SOMA
survey. K.E.I.T. acknowledges support from NAOJ ALMA Scientific Research grant number 2017$-$05A. This research made use of APLpy, an open-source
plotting package for Python hosted at
http://aplpy.github.com. We thank the anonymous referee whose comments improved this manuscript.

\software{CASA \citep{2007ASPC..376..127M}, APLpy \citep{2012ascl.soft08017R}
, HOCHUNK3d \citep{2003ApJ...591.1049W,2013ApJS..207...30W}, CLOUDY \citep{2013RMxAA..49..137F}}

\bibliography{SOMA_biblio}

\end{document}